\documentclass[fleqn,usenatbib]{mnras}

\usepackage{newtxtext,newtxmath}

\usepackage[T1]{fontenc}

\DeclareRobustCommand{\VAN}[3]{#2}
\let\VANthebibliography\thebibliography
\def\thebibliography{\DeclareRobustCommand{\VAN}[3]{##3}\VANthebibliography}

\usepackage{graphicx}	
\usepackage{amsmath}	
\usepackage{adjustbox}
\usepackage{subfigure}
\usepackage{subcaption}
\usepackage{caption}
\usepackage{multirow}
\usepackage{graphics}
\usepackage{comment}

\newcommand{\msun}{\,M$_{\odot}$}
\newcommand{\mhe}{\,M$_{\mathrm{He,i}}$}
\newcommand{\mns}{\,M$_{\mathrm{NS}}$}
\newcommand{\mco}{\,M$_{\mathrm{CO}}$}

\title[Heavy DNSs]{Formation of heavy double neutron stars I: Eddington-limited accretion for a 1.4\,\msun{} neutron star at solar metallicity}

\author[Nair et al.]{
Ashwathi Nair$^{1,2}$\thanks{E-mail: ashwathinair@swin.edu.au}
and
Simon Stevenson$^{1,2}$
\\
$^{1}$OzGrav : The ARC Centre of Excellence for Gravitational Wave Discovery, Hawthorn, VIC 3122, Australia\\
$^{2}$Centre for Astrophysics and Supercomputing, Swinburne University of Technology, Hawthorn, VIC 3122, Australia\\
}

\date{Accepted XXX. Received YYY; in original form ZZZ}

\pubyear{\the\year{}}

\begin{document}
\label{firstpage}
\pagerange{\pageref{firstpage}--\pageref{lastpage}}
\maketitle

\begin{abstract}
More than 30 Galactic double neutron star (DNS) binaries have now been identified through radio pulsar timing.
The 24 DNSs in the Galactic field with measured total masses lie in the narrow range of $2.3$--$2.9$\,\msun{}.
In contrast, gravitational-wave observations have detected two DNS mergers: GW170817, with a total mass of $2.7$\,\msun{}, and GW190425, with a significantly higher mass of $3.4$\,\msun{}. 
The unusually high mass of GW190425 suggests a non-standard formation channel not represented in the known Galactic population. To investigate the origin of such a massive DNS system, we model the late evolutionary stages of helium stars with initial masses between $2.5$ and $9.8$\,\msun{} in binaries with $1.4$\,\msun{} neutron star companions, using the 1D stellar evolution code MESA at solar metallicity. We test alternative formation pathways and calibrate our models to reproduce the observed Galactic DNS mass and orbital distributions. 
By incorporating a modified natal kick prescription, our population synthesis results are broadly consistent with the observed total mass distribution of known DNS systems. Only a small fraction of DNSs of our model have total masses $\ge3$\,\msun{}, insufficient to explain the high rate of massive DNS mergers inferred from GW observations. However, our model rules out the formation of heavy DNS systems like GW190425 via the second unstable mass transfer.
\end{abstract}

\begin{keywords}
gravitational waves -- stars: massive -- stars: evolution -- binary evolution
\end{keywords}



\section{Introduction}
\label{sec:intro}

The evolution of massive stars, the formation of double neutron star (DNS) systems, and the outcomes of neutron star mergers are central to several domains of modern astrophysics. 
Variations in key physical processes, such as mass transfer episodes, orbital evolution, and the final structure of the helium star, including its metal core and remaining envelope, can lead to different supernova (SN) explosion types and provide insight into their underlying mechanisms. \citep{1973ApJ...186.1007W, 1984ApJS...54..335I, Yoon_2010, Tauris:2013ApJL}. 
The merging of neutron stars may give rise to short gamma-ray bursts \citep[GRB;][]{Narayan:1992ApJ}, accompanied by optical transients known as kilonovae \citep[][]{LattimerSchramm:1974ApJL,1998ApJ...507L..59L}. 
Extreme physical conditions in kilonovae produce heavy r-process elements \citep[e.g.,][]{2015IJMPD..2430012R, 2015MNRAS.448..541J}. 
Understanding GRBs and kilonovae will enhance our understanding of the origin of heavy elements in the universe, particularly those formed through r-process nucleosynthesis following neutron star mergers.

DNSs are also observable as radio pulsars for millions of years, and they eventually merge due to gravitational wave (GW) radiation. These emissions are powerful and are detected by gravitational wave detectors such as LIGO \citep[][]{2015CQGra..32g4001L}, Virgo \citep[][]{2015CQGra..32b4001A}, and KAGRA \citep[][]{2019NatAs...3...35K}. The entire binary evolution of these massive stars goes through several complex physical processes. These systems undergo both stable and unstable mass transfer, as well as two supernova explosions (SNe). 
Systems that survive these critical stages can form DNS binaries.
As such, DNS systems serve as valuable tracers of binary evolutionary pathways, offering key insights into the formation, evolution, and fate of massive binary stars.


Within the scope of our research, we have embarked upon an extensive examination of DNS systems across two distinct spectra: Galactic radio sources and GW sources of extragalactic origin.  
Radio observations have unveiled numerous binary neutron star systems. Recently, the rate of discovering new Galactic DNS systems has increased, thanks to new high-sensitivity instruments such as MeerKAT and the Five-hundred-meter Aperture Spherical radio Telescope \citep[FAST;][]{2011IJMPD..20..989N}. 
We have identified a total of approximately 34 Galactic DNS systems \citep[e.g.,][See Table~\ref{tab:dns_candidates} for the updated list of Galactic DNS]{Tauris:2017ApJ, Sengar_2022, Agazie_2021, Andrews:2019ApJ, Swiggum_2023, 2023ApJ...958L..17W, 2024ApJ...964L...7Z}. These systems exhibit total masses ranging from $2.30$ to $3.88$\,\msun{} \citep[if the massive pulsar binary PSR $J0514$-$4002$E is a DNS;][]{Barr_2024}. The 24 confirmed DNS systems in the field typically fall in the narrower range of $2.3$--$2.9$\,\msun{}.

However, GW observations have yielded two merger detections of DNS systems so far: GW170817 with component masses between $0.86$--$2.26$\,\msun{} and a combined mass of $2.74^{+0.04}_{-0.01}$\,\msun{} \citep{Abbott:2017PhRvL}, being the only GW event associated with an electromagnetic counterpart \citep{Abbott:2017ApJLMultimessenger,Abbott_2017_d,Abbott_2017_c, Andreoni_2017}, and GW190425 with component masses between $1.12$--$2.52$\msun{} and a combined mass of $3.4^{+0.3}_{-0.1}$\msun{} \citep[][]{Abbott:2020ApJL}.
The Fourth LIGO/Virgo/KAGRA (LVK) observing run (O4) is currently underway, but so far, no new significant DNS candidates have been reported in low-latency.

\begin{table*}
\centering
\caption{Galactic Double Neutron Star Candidates, in order of ascending orbital period $P_\mathrm{b}$. 
All systems are found in the Galactic field, except for those listed under the Globular Cluster heading. 
\textbf{References:} Confirmed DNS:
(1) \protect\citet{2018ApJ...854L..22S},
(2) \protect\citet{PhysRevX.11.041050},
(3) \protect\citet{Cameron:2018MNRAS},
(4) \protect\citet{2020Natur.583..211F},
(5) \protect\citet{2014MNRAS.443.2183F},
(6) \protect\citet{2016ApJ...829...55W},
(7) \protect\citet{2018ApJ...859...93L},
(8) \protect\citet{Fonseca_2014},
(9) \protect\citet{2024ApJ...964L...7Z},
(10) \protect\citet{2023A&A...678A.187C},
(11) \protect\citet{2021MNRAS.500.4620H},
(12) \protect\citet{Sengar_2022},
(13) \protect\citet{Agazie_2021},
(14) \protect\citet{2017ApJ...851L..29M},
(15) \protect\citet{Padmanabh_2023},
(16) \protect\citet{2015ApJ...812..143M},
(17) \protect\citet{Janssen_2008},
(18) \protect\citet{Swiggum_2023},
(19) \protect\citet{2009MNRAS.393..623K},
(20) \protect\citet{Su_2024},
(21) \protect\citet{refId0},
(22) \protect\citet{Swiggum_2015},
(23) \protect\citet{Wang:2025RAA}; 
Possible DNS:
(24) \protect\citet{2015ApJ...798..118V},
(25) \protect\citet{Ng:2018MNRAS},
(26) \protect\citet{2023ApJ...958L..17W};
Globular Cluster DNS:
(27) \protect\citet{1989Natur.337..531W,Jacoby:2006ApJL},
(28) \protect\citet{2023ApJ...942L..35B},
(29) \protect\citet{Barr_2024},
(30) \protect\citet{2015ApJ...807L..23D},
(31) \protect\citet{Lynch:2012ApJL},
(32) \protect\citet{2019MNRAS.490.3860R},
(33) \protect\citet{Padmanabh_2024}.
}

\label{tab:dns_candidates}
\begin{adjustbox}{max width=\textwidth}
\begin{tabular}{lcccccccccc}
    \hline
    \hline
    PSR & $P_0$ & $\dot{P}$ & $P_\textrm{b}$ & $e$ & $M_\textrm{t}$ & $M_\textrm{c}$ & $M_\textrm{p}$ & $\tau_\textrm{m}$ & Merges within \\ 
     & (ms) & (s\,s$^{-1}$) & (days) & & (M$_\odot$) & (M$_\odot$) & (M$_\odot$) & (Gyr) & Hubble time\\
    \hline
    \hline
    \multicolumn{10}{c}{\textbf{Confirmed DNS}} \\
    \hline
     J1946+2052$^1$ & 16.960 & 9.20$\times10^{-19}$ & 0.078 & 0.064 & 2.50(4) & $>$1.18 & $<$1.31 & \textasciitilde0.0455\,& Yes\\
     J0737$-$3039A$^2$ & 22.699 & 1.76$\times10^{-18}$ & 0.102 & 0.088 & 2.587052(9) & 1.248868(13) & 1.338185(14) & 0.0860& Yes\\
     J0737$-$3039B$^2$ & 2773.5 & 8.92$\times10^{-16}$ & 0.102 & 0.088 & 2.587052(9) & 1.338185(14) & 1.248868(13) & 0.0860& Yes\\
     J1757$-$1854$^3$ & 21.497 & 2.63$\times10^{-18}$ & 0.184 & 0.606 & 2.732876(8) & 1.3917(4) & 1.3412(4) & 0.0761& Yes\\
     J1913+1102$^4$ & 27.285 & 1.61$\times10^{-19}$ & 0.206 & 0.090 & 2.8887(6) & 1.27(3) & 1.62(3) & 0.470& Yes\\ 
     J1756$-$2251$^5$ & 28.462 & 1.02$\times10^{-18}$ & 0.32 & 0.181 & 2.56999(6) & 1.230(7) & 1.341(7) & 1.66& Yes \\ 
     B1913+16$^6$ & 59.030 & 8.62$\times10^{-18}$ & 0.323 & 0.617 & 2.828(1) & 1.390(1) & 1.438(1) & 0.301& Yes\\ 
     J0509+3801$^7$ & 76.541 & 7.93$\times10^{-18}$ & 0.38 & 0.586 & 2.805(3) & 1.46(8) & 1.34(8) & 0.576& Yes \\ 
     B1534+12$^8$ & 37.904 & 2.42$\times10^{-18}$ & 0.421 & 0.274 & 2.678463(4) & 1.3455(2) & 1.3330(2) & 2.73& Yes \\
     J1846$-$0513$^9$& 23.369 & $1.0106(3) \times 10^{-18}$ & 0.61302 & 0.208(9) & 2.6287(35) & $>$1.28 & $<$1.35 & & \\
     J1208$-$5936$^{10}$& 28.714 & $< 4 \times 10^{-20}$ & 0.632 &  0.348 & 2.586(6) & ${1.32^{+0.25}_{-0.13}}$ & ${1.26^{+0.13}_{-0.25}}$& 7.2(2)& Yes \\
     J1829+2456$^{11}$ & 41.010 & 5.25$\times10^{-20}$ & 1.176 & 0.139 & 2.60551(19) & 1.299(4) & 1.306(7) & 55 & No\\
     J1325$-$6253$^{12}$ & 28.969 & 4.80$\times10^{-20}$ & 1.816 & 0.064 & 2.57(6) & $>$0.98 & $<$1.59 & \textasciitilde189\, &No\\
     J1759+5036$^{13}$ & 176.02 & 2.43$\times10^{-19}$ & 2.043 &  0.308 & 2.62(3) & $>$0.7006 & $<$1.9194 & \textasciitilde177\, &No\\
     J1411+2551$^{14}$ & 62.453 & 9.56$\times10^{-20}$ & 2.616 &  0.17 & 2.538(22) & $>$0.92 & $<$1.62 & \textasciitilde466\,&No \\
     J1155-6529$^{15}$ & 78.88  & $\sim$3.5$\times10^{-19}$  & 3.67 & 0.26 & ... & $>1.27$ & 1.4 & & \\
     J0453+1559$^{16}$ & 45.782 & 1.86$\times10^{-19}$ & 4.072 & 0.113 & 2.734(4) & 1.174(4) & 1.559(5) & 1\,453& No \\ 
     J1518+4904$^{17}$ & 40.935 & 2.72$\times10^{-20}$ & 8.634 & 0.249 & 2.7183(7) & 1.31(8) & 1.41(8) & 8\,844& No \\
     J1018$-$1523$^{18}$ & 83.152 & 1.09(6)$\times10^{-19}$ & 8.984 &  0.228 & 2.3(3) & $>$1.16 & $<$1.1(3) & ~\textasciitilde1.4(3)$\times10^4$\,& No \\
     J1753$-$2240$^{19}$ & 95.138 & 9.70$\times10^{-19}$ & 13.638 &  0.304 & ... & $>$0.4875 & ... & ...\,& No \\
     J1901+0658 $^{20}$& 75.7 & $2.169(6)\times 10^{-19}$ & 14.45477 & 0.3662392  & 2.79(7) & $>$ 1.11 & $<$ 1.68 & $ > 10^{3}$ & No\\[0.3ex]
     J1811$-$1736$^{21}$ & 104.18 & 9.01$\times10^{-19}$ & 18.779 & 0.828 & 2.57(10) & $>$0.93 & $<$1.64 & \textasciitilde1\,800 & No\\ 
     J1930$-$1852$^{22}$ & 185.52 & 1.80$\times10^{-17}$ & 45.06 &  0.399 & 2.59(4) & $>$1.30 & $<$1.25&&No\\[0.3ex]
     J0528+3529 $^{23}$ & 78.2 & $7.36 \times 10^{-19}$ & 11.72618 & 0.2901088(10) & 2.90 & $>$1.2 & $<$1.8& ... & ... \\ 
     J1844-0128 $^{23}$ & 29.1 & $3.1 \times 10^{-20}$ & 20.96779 & 	0.234949(5) & ... & >0.74 &1.4 & ...& ...\\
    \hline
    \hline
    \multicolumn{10}{c}{\textbf{Possible DNS}} \\
    \hline
    J1906+0746$^{24}$ & 144.07 & 2.03$\times10^{-14}$ & 0.166 &  0.085 & 2.6134(3) & 1.322(11) & 1.291(11) & 0.308& Yes \\ 
    J1755$-$2550$^{25}$ & 315.20 & 2.43$\times10^{-15}$ & 9.696 &  0.089 & ... & >0.39 & ... & ...\, & ...\\ 
     J2150+3427$^{26}$& 654 & $3.60(4)\times 10^{-18}$ & 10.59213  & 0.601494(2) & 2.59(13) &  $> 0.98$& $< 1.67$ & ... & ... \\
    \hline
    \hline
    \multicolumn{10}{c}{\textbf{Globular Clusters}} \\
    \hline
    J2127+11c$^{27}$ & 30.53 & ... & 0.335 & 0.681 & 2.712(13) & 1.354 & 1.358(10) & ... & ...\\
    J2140-2311B$^{28}$ & 13 & $-$ 6(25) $\times 10^{19}$ & 6.215 & 0.879 & 2.53(8) & $> 1.10$ & $<1.43$ & ... & ...\\
    J0514-4002E$^{29}$ & 5.5959 & ... & 7.447 &  0.707 & 3.8810 & 2.09 & 1.79 & ... & ...\\
    J1835-3259A$^{30}$ & 3.89 & ... & 9.246 & 0.968 & ... & $> 0.76$ & 1.4 & ... & ...\\
    J1807-2500B$^{31}$ & 4.2 & ... & 9.956 & 0.74703 & 2.571(73) & 1.2064 & 1.3655 & ... & ...\\
    J0514-4002A$^{32}$ & 4.99 & ... & 18.785  & 0.8879 & 2.4730(6) & 1.22(+6/-5) & 1.25 (+5/-6) & ... & ...\\
    J1748-2446ao$^{33}$ & 2.27 & ... &57.555  &  0.324 & 3.166 & $> 0.93$ & $< 2.23$ & ... & ...\\    
    \hline
\end{tabular}
\end{adjustbox}

\end{table*}

We expect to develop an improved picture by combining radio and GW observation data to understand the DNS population. However, a comparison of their observed properties suggests that the radio and GW-detected DNS populations may differ significantly. GW170817 displays component masses and a combined mass consistent with Galactic DNS systems. While GW190425 exhibits component masses pretty well in agreement with the masses of NSs found in binaries, the observed final mass is significantly higher than any known DNS remnants, hinting at alternative formation pathways for such heavy DNS systems. Two well-known primary formation pathways for DNS mergers are the `common envelope isolated binary evolution channel', where two neutron stars form sequentially through supernovae (SNe) events within an isolated binary system, and the `dynamical formation channel' where binary systems originate within densely populated stellar environments, such as globular star clusters.
  
Theoretical investigations into DNS formation have been conducted for decades. The commonly accepted canonical isolated DNS formation pathway has emerged (e.g. \citealt{1991PhR...203....1B} and \citealt{2006csxs.book..623T}), which is briefly summarized and presented in \cite{Tauris:2017ApJ}. 
The primary massive OB-type star evolving from the Zero-Age Main Sequence (ZAMS) undergoes mass transfer, shedding the hydrogen envelope, and evolves into a naked helium (He) star.
These binaries consisting of subdwarf OB stars with massive B/Be type companions have recently been observed \citep[][]{Drout:2023Sci}.
The stripped helium star then undergoes a SN explosion, leaving behind a neutron star with the secondary OB star companion. 
The binary system is observable at this stage as a high-mass X-ray binary (HMXB) system. When the secondary star starts expanding, a crucial phase in the evolutionary pathway to DNS formation occurs, where a giant star fills its Roche lobe and, depending on the mass ratio, initiates a dynamically unstable Case B Roche-lobe Overflow (RLO) mass-transfer (MT) episode onto the NS companion. The stellar core and the NS become engulfed by the expanding envelope, orbiting each other inside the envelope. This process is known as Common Envelope (CE) evolution \citep{1976IAUS...73...75P, 1993PASP..105.1373I, 2013Sci...339..433I}. Here, the process of dynamical friction during the motion of a neutron star (NS) within the envelope of a massive star frequently leads to a significant reduction in the system's orbital angular momentum and, in certain instances, results in the ejection of the hydrogen-rich envelope. 
This leads to the formation of a compact binary containing a helium-rich star with a reduced mass of a few solar masses. This stripped helium star typically resides in a close, nearly circular orbit with the companion NS. 
Subsequent expansion of the helium star following core helium burning prompts an additional phase of stable Case BB RLO MT within the binary system for low-mass helium stars \citep[e.g.,][]{Tauris:2013ApJL,Tauris:2015MNRAS}. During this transfer, mass is exchanged from the helium-rich donor star, initiating the recycling of the pulsar.
In the tightest binaries, much of the helium envelope of the donor star is removed, leaving behind an ultra-stripped metal core \citep[][]{Tauris:2013ApJL,Tauris:2015MNRAS}. 
Finally, the metal core undergoes a supernova and produces a second neutron star. If the binary remains compact, it gives birth to a DNS system. 
The systems formed via this channel give rise to DNS systems such as GW170817 \citep{Abbott:2017PhRvL} and those observable within our Milky Way galaxy \citep[][]{Tauris:2017ApJ}.
  
An ongoing question concerns the absence of heavy DNS systems in radio observations. One hypothesis suggests that these massive DNS systems may merge rapidly, known as `fast-mergers', eluding radio detection. 
\citet{Abbott:2020ApJL} introduce a formation pathway for such fast-merging systems in an attempt to explain the origin of GW190425 \citep[see also][]{2020MNRAS.496L..64R}.
This channel involves two neutron stars with differing birth histories, undergoing unstable mass transfer \citep{2003ApJ...592..475I, Belczynski_2002, Vigna-Gomez:2018MNRAS}, leading to compact, sub-hour orbital period binaries. If the binary survives the common envelope phase, the high mass can be resultant of this formation channel. \citet{Vigna_G_mez_2021} introduced an alternate formation channel in which a massive helium star of mass $\geq 9$\,\msun{}, due to its limited radial expansion, avoids transferring mass onto its companion neutron star, avoiding recycling the pulsar.
After a few tens of millions of years, non-recycled pulsars become radio-quiet, such that these binary systems can only be detected by gravitational waves, which hints at why such heavy DNSs might not be observable at radio wavelength. 

On the other hand, \citet{Galaudage_2021} claim that GW190425 may not be a statistical outlier after all, and may be consistent with the Galactic DNS mass distribution. 
Indeed, population synthesis models often predict the formation of massive DNSs like GW190425 at solar metallicity \citep[e.g.,][]{Kruckow:2020A&A,Mandel:2021MNRAS}, despite the fact that no such massive DNSs are observed in the Galactic population. 
\citet{2018MNRAS.480.2011G} find that massive DNSs tend to form at lower metallicities. 
\citet{Zhang:2023MNRAS} propose super-Eddington accretion as a possible scenario to form GW190425. 
\citet{Qin:2024A&A} propose the formation of massive GW190425-like DNSs through stable Case BB MT with a massive first-born neutron star.
These uncertainties in the formation of GW190425 force us to revisit the physical processes that lead to the formation of such massive DNS systems.

One approach to investigate DNS system properties is the use of binary population synthesis. It can broadly reproduce most of the observed characteristics of the Galactic DNS population \citep[for e.g.,][]{Andrews:2015ApJ,2018MNRAS.481.1908K,2018ApJ...867..124S,Vigna-Gomez:2018MNRAS,Chattopadhyay:2020MNRAS}. 
However, the properties of DNS populations modeled with binary population synthesis are typically unable to match the orbital properties of DNSs (orbital period and eccentricity) in detail, as they often produce too many intermediate and high eccentricity DNSs \citep[$e = 0.4$--$0.6$;][]{Ihm_2006,Andrews:2015ApJ, 2017AcA....67...37C}, which is not the case for the observed Galactic systems \citep[][]{Beniamini:2016MNRAS,Andrews:2019ApJ}. A similar situation exists for Galactic Be X-ray binaries \citep[][]{Valli:2025arXiv}.
Additionally, population synthesis models typically overpredict the masses of Galactic DNS systems \citep[e.g.,][]{Vigna-Gomez:2018MNRAS,Chattopadhyay:2020MNRAS,Mandel:2021MNRAS}.

Given these open questions about DNSs, in this work, we aim to calibrate the models by ensuring that they can explain the Galactic DNS population. In this work, we do not address the DNS formed dynamically, i.e, in a globular cluster. In Section~\ref{sec:model_calculation}, we briefly explain the stellar evolution code used, details about the initial conditions for binary evolution of helium-neutron star systems, and the role of helium star evolution in DNSs evolutionary process. We discuss the standard and modified SN and kick prescriptions we use to post-process the post-SN binary systems and the effects of SN on the post-SN orbital dynamics. In section~\ref{sec:Results}, we discuss the results of the population of He-NS binaries and examine the post-SN orbital properties like the orbital period '$P_{\mathrm{orb}}$', eccentricity 'e', and its distribution along with the DNS total mass distribution. We also briefly discuss the progenitors of GW190425. Finally we conclude in section~\ref{sec:conclusions}.

\section{Detailed Binary Models}
\label{sec:model_calculation}

We perform detailed binary evolution using the one-dimensional (1D) stellar evolution code MESA \citep[][]{Paxton2011,Paxton2013,Paxton2015,Paxton2018,Paxton2019,Jermyn2023}, version \texttt{23.05.01}.
We start by modeling helium stars in detail within the mass range of $2.5$--$10$\,\msun{} with a helium mass fraction $Y=0.98$, at solar metallicity $Z=0.02$.
We then let our He-stars evolve through the He-ZAMS stage and gradually to the fast phase of helium shell burning, where the model has obtained its maximum radius, using the MESA module \texttt{MESAstar}. 
We use the 'Dutch' prescription for the mass lost by hot winds (\citealt{2000A&A...360..227N}; \citealt{2001A&A...369..574V}; \citealt{2009A&A...497..255G}),
with a scaling factor of 0.8 for massive helium stars (\citealt{2000A&A...361..101M}). 
We adopt the Ledoux criterion to treat the boundaries of the convective zones and model convection using the mixing length alpha $\alpha_{ml} = 2$ and alpha semiconvection $\alpha_{sc} = 1$. 
Gold tolerances for intermediate and massive helium star models are used.
We use Type2 Rosseland mean opacity table given by \citet{1996ApJ...464..943I} for $Z = 0.02$. 

We then pair the simulated single ZAMS helium star models with a $1.4$\,\msun{} NS companion to obtain a detailed evolutionary model of a DNS progenitor system. 
The starting point of the evolution in this simulation is the post-CE phase in the standard DNS formation pathway. 
If the hydrogen envelope is successfully ejected, the binary survives and leaves behind a naked, stripped helium star with a firstborn NS companion. 
Recent models have shown that following the common envelope phase, a small amount of hydrogen ($\lesssim 1$\,M$_\odot$) may be retained on the surface of the helium star \citep[][]{Nie:2025ApJ}. In this study we neglect this hydrogen layer and assume that common envelope evolution leaves behind naked helium stars.

The binaries are evolved using the MESA module \texttt{Mesabinary}. 
The binary is initially set up in a circular orbit $(e = 0)$.
To compute the mass transfer rate from the He star $|\dot M_\mathrm{{He}}|$, we use the 'Ritter' mass transfer scheme (\citealt{1988A&A...202...93R}). 
We assume that the mass accretion rate for helium rich materical onto a neutron star is limited to the Eddington accretion rate, $\dot{M}_\mathrm{Edd} = 3 \times 10^{-8}$\,M$_{\odot}$\,yr$^{-1}$.
The mass transfer from the helium donor star is typically orders of magnitude higher than this, leading to highly non-conservative mass transfer \citep[e.g.,][]{Tauris:2015MNRAS}.
For simplicity, we therefore assume completely non-conservative mass transfer, that is the fraction of mass lost from the vicinity of the NS in the form of a fast wind is set to be 1.0.
Orbital angular momentum loss due to mass loss assumes that mass is re-emitted isotropically from the accretor \citep{1997A&A...327..620S}. 
Orbital angular momentum loss due to gravitational wave radiation is included, but is mostly negligible.

\subsection{Role of Helium Stars in late Double Neutron Star merger formation}
\label{subsec:He_star_in_DNS}

\begin{figure}
    \centering
    \includegraphics[]{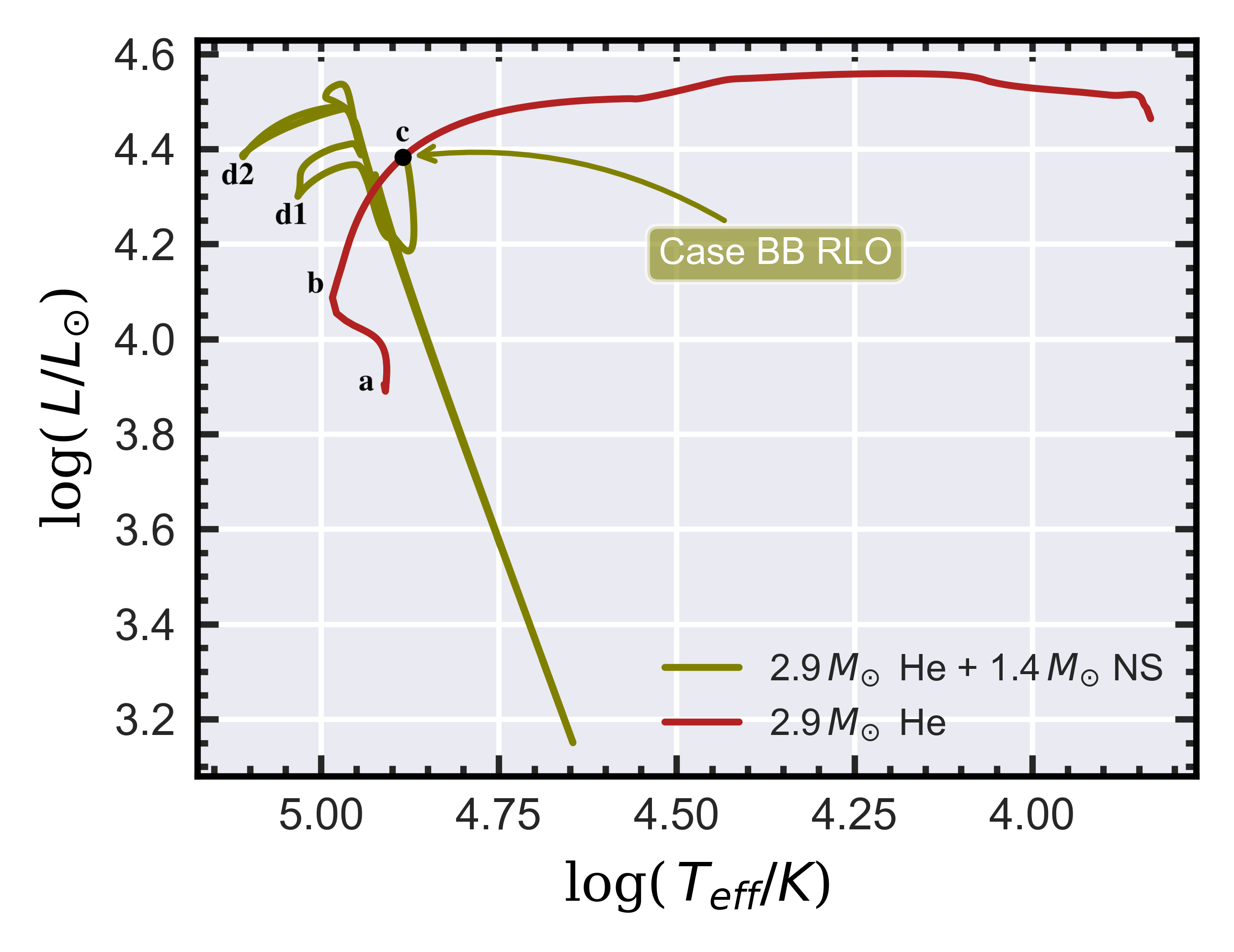}
    \caption{The Hertzsprung-Russel (HR) diagram of the evolution of 
 a $2.9\,$\msun{} He + 1.4\,\msun{} NS binary system (olive green solid line) compared to the evolution of a single 2.9\,\msun{} He star (maroon solid line). The points a-d2 indicate different evolutionary phases of the He-NS binary (see text in section~\ref{subsec:He_star_in_DNS} for a detailed description).}
    \label{fig:HRD}
\end{figure}

At the starting point of the post-common envelope phase, we assume the first-born neutron star to have a standard mass of $1.4$\,\msun{}. 
Instead of solving the stellar structure equations of NSs, they are assumed to be a point mass. 
Fig~\ref{fig:HRD} shows the evolutionary track of a single $2.9$\,\msun{} He-ZAMS star on the Hertzsprung-Russell (HR) diagram, along with a tight (initial orbital period $P_{\mathrm{orb, i}}$ of $0.145$\,d), $2.9$\,\msun{} He-ZAMS star -- $1.4$\,\msun{} NS binary.
The points indicated in the figure correspond to the following: (a) helium star ZAMS: the start of core helium burning; (b) the core helium is exhausted. The star begins helium shell burning and starts to expand, reaching a luminosity of approximately 
 $\mathrm{log} (L/L_{\odot}) \simeq 4.1$, (c) onset of Case BB RLO - The star fills its Roche lobe during the helium shell burning phase and begins transferring mass to the neutron star companion; (d1) After mass transfer, the star begins core carbon burning, which results in contraction of the star’s radius.; (d2) consecutive ignitions of carbon-burning shells.

\begin{figure}
    \centering
    \includegraphics[]{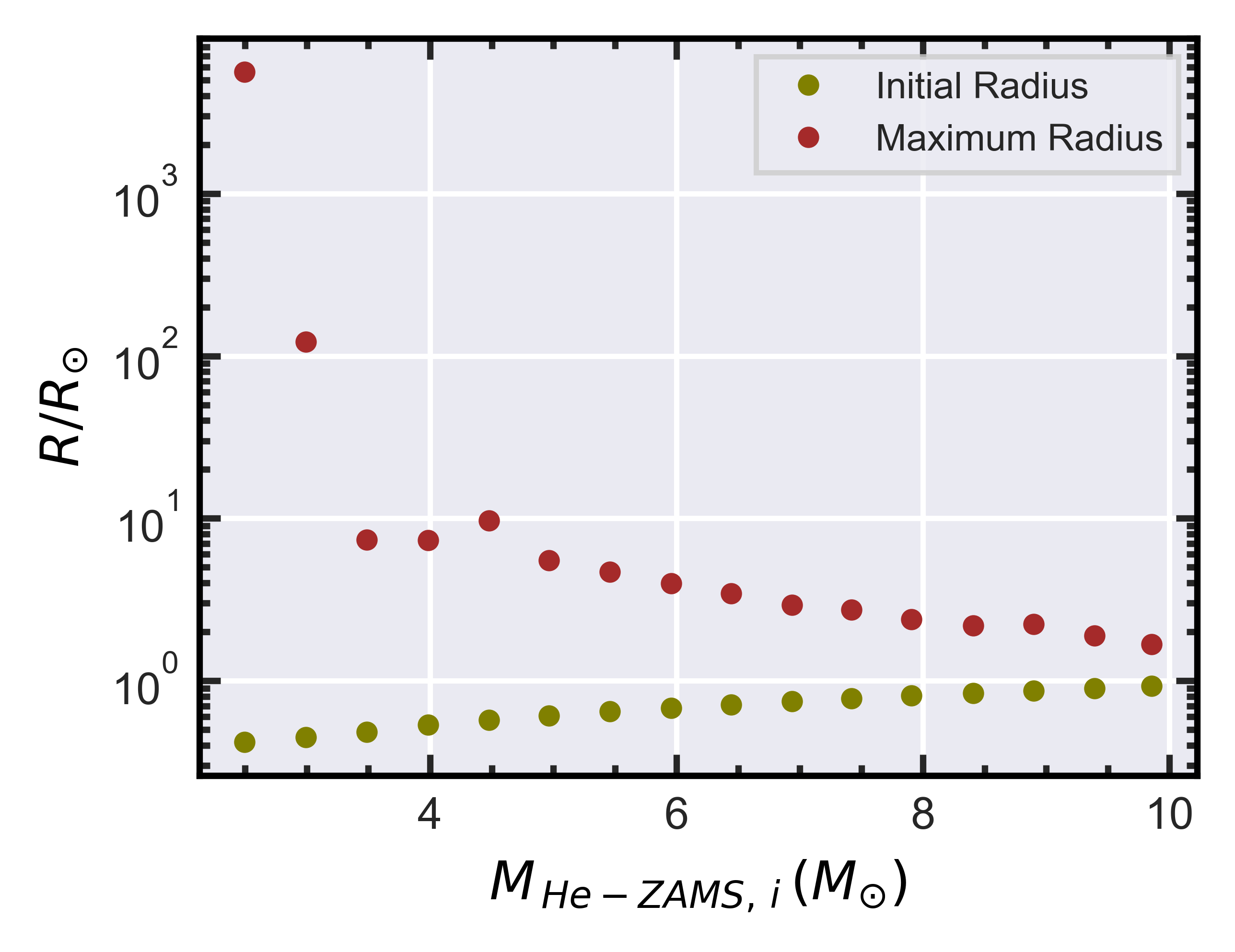}
    \caption{Initial (olive green solid circles) and maximum radius (maroon solid circles) of single He-ZAMS stars as a function of their initial ZAMS masses ranging between $2.5$--$9.8$\,\msun{}}
    \label{fig:rad_as_fun_of_ini_mass}
\end{figure}

Fig~\ref{fig:rad_as_fun_of_ini_mass} shows the initial and maximum radii of our helium star models. We can see that helium stars with ZAMS mass $2.5$--$2.9$\,\msun{} expand to significantly larger dimensions than He-ZAMS with masses $\geq 3.4$\,\msun{}. The substantial radial expansion of a $2.9$\,\msun{} He star in a binary will lead it to overfill its Roche lobe and lose considerable mass by mass transfer, which is a key factor for Case BB RLO and DNS systems. Helium stars more massive than 5\,\msun{} expand by less than a factor of 10 compared to their ZAMS radius, meaning that they will only overflow their Roche lobe in binary systems with a narrow range of separations.

\begin{figure*}
\subfigure[]
{\includegraphics[]{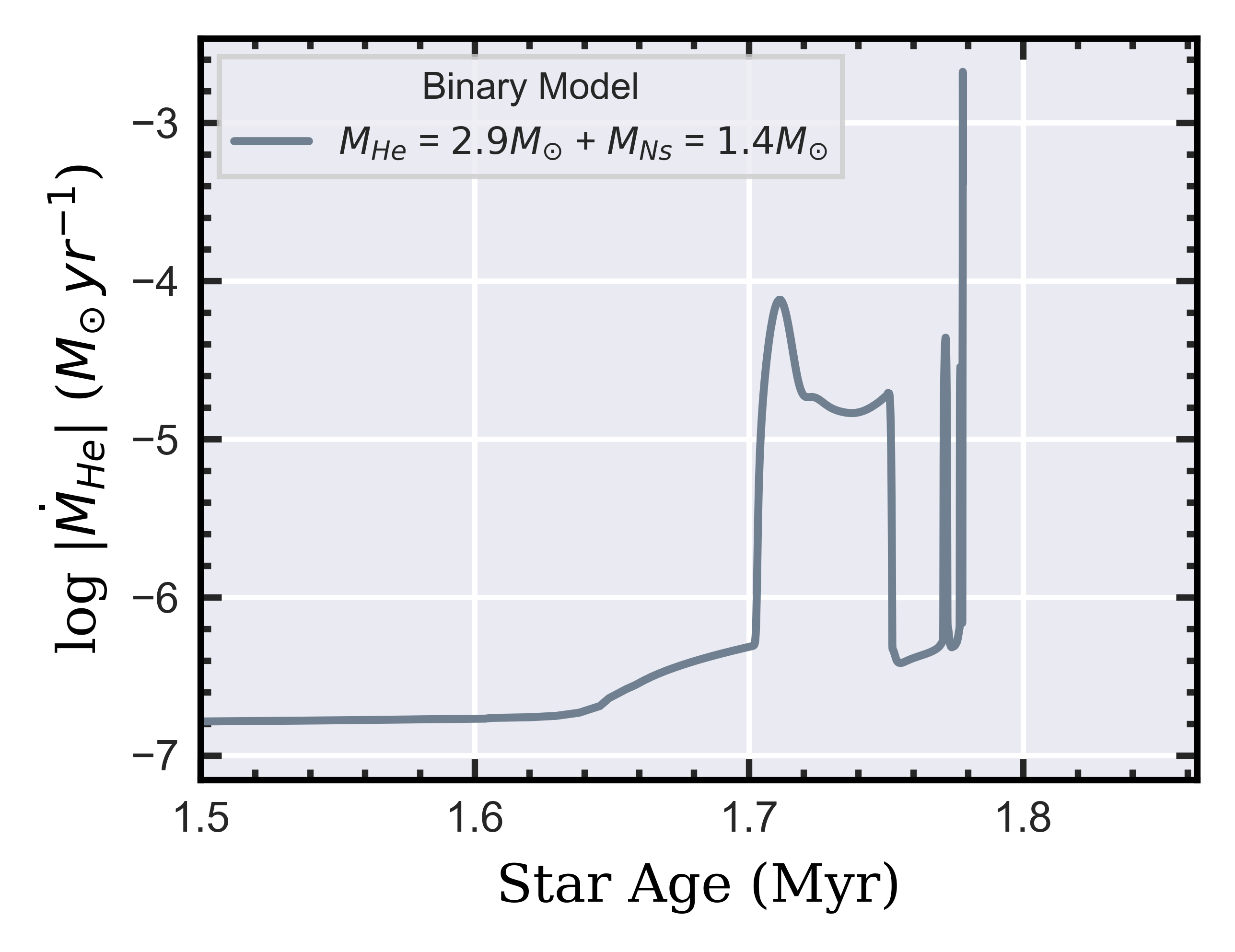}\label{fig:mass_transfer}}
\hspace{0.5cm}
\subfigure[]
{\includegraphics[]{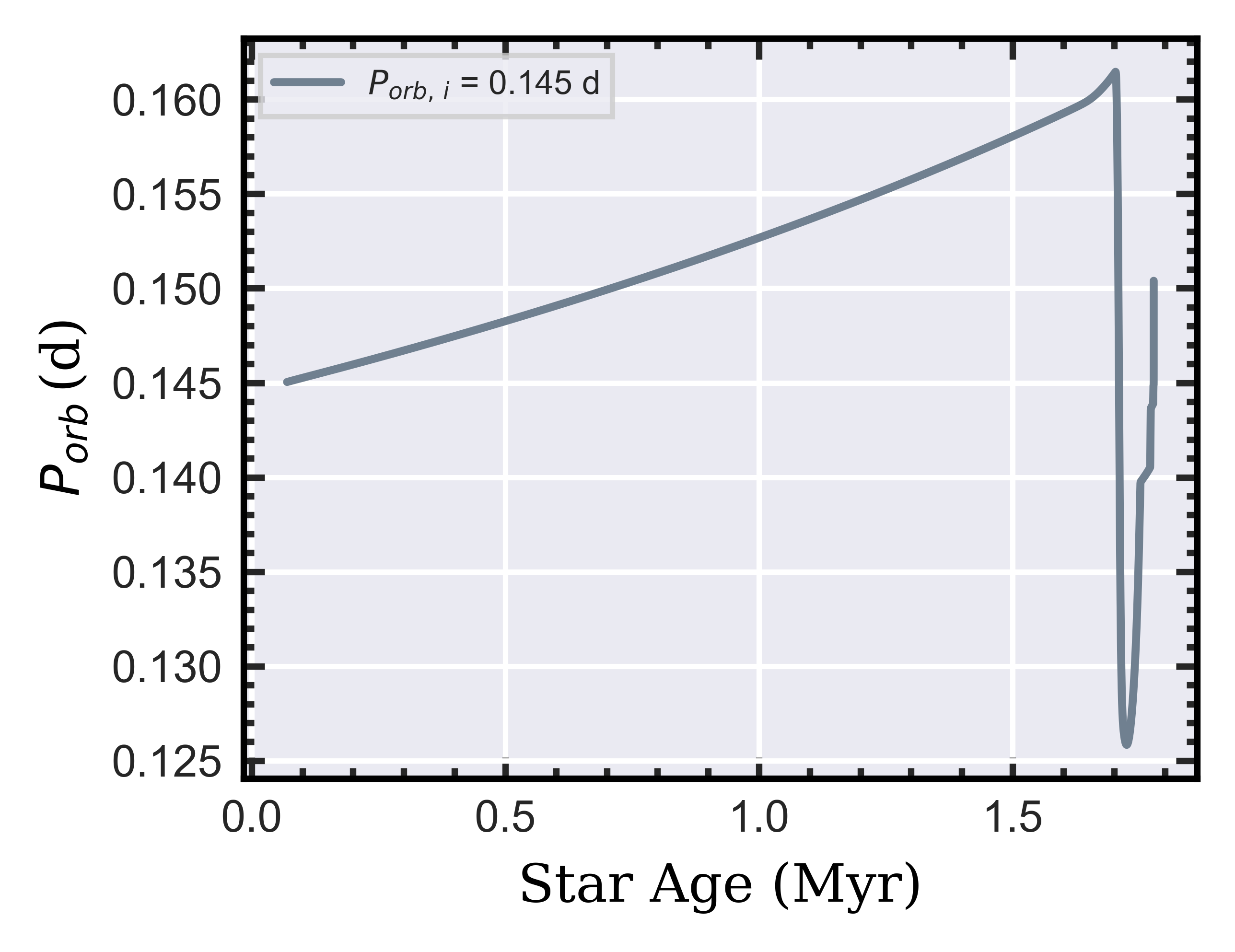}\label{fig:2.9_orbital_period}}
\hspace{0.5cm}

\caption{(Left) Mass transfer rate, and (right) orbital period as a function of stellar age for a 2.9\,\msun{} He + 1.4\,\msun{} NS binary. }
 \label{fig:mass_transfer_orbital_period}
\end{figure*}

Fig~\ref{fig:mass_transfer} shows the mass transfer rate as a function of time for the 2.9\,\msun{} He star--1.4\,\msun{} NS binary discussed earlier. Case BB RLO is initiated at 1.71 Myr. Before the onset of RLO, the wind mass loss from the He star gradually increases from $10^{-6.95}$\,\msun{}\,$yr^{-1}$ to $10^{-6.70}$\,\msun{}\,$yr^{-1}$, thereby increasing $P_{\mathrm{orb}}$ from 0.145\,d to 0.16\,d and drops suddenly as the companion helium star starts transferring mass onto the NS (see fig~\ref{fig:2.9_orbital_period}), thereby recycling it. 

\begin{figure}
    \centering
    \includegraphics[]{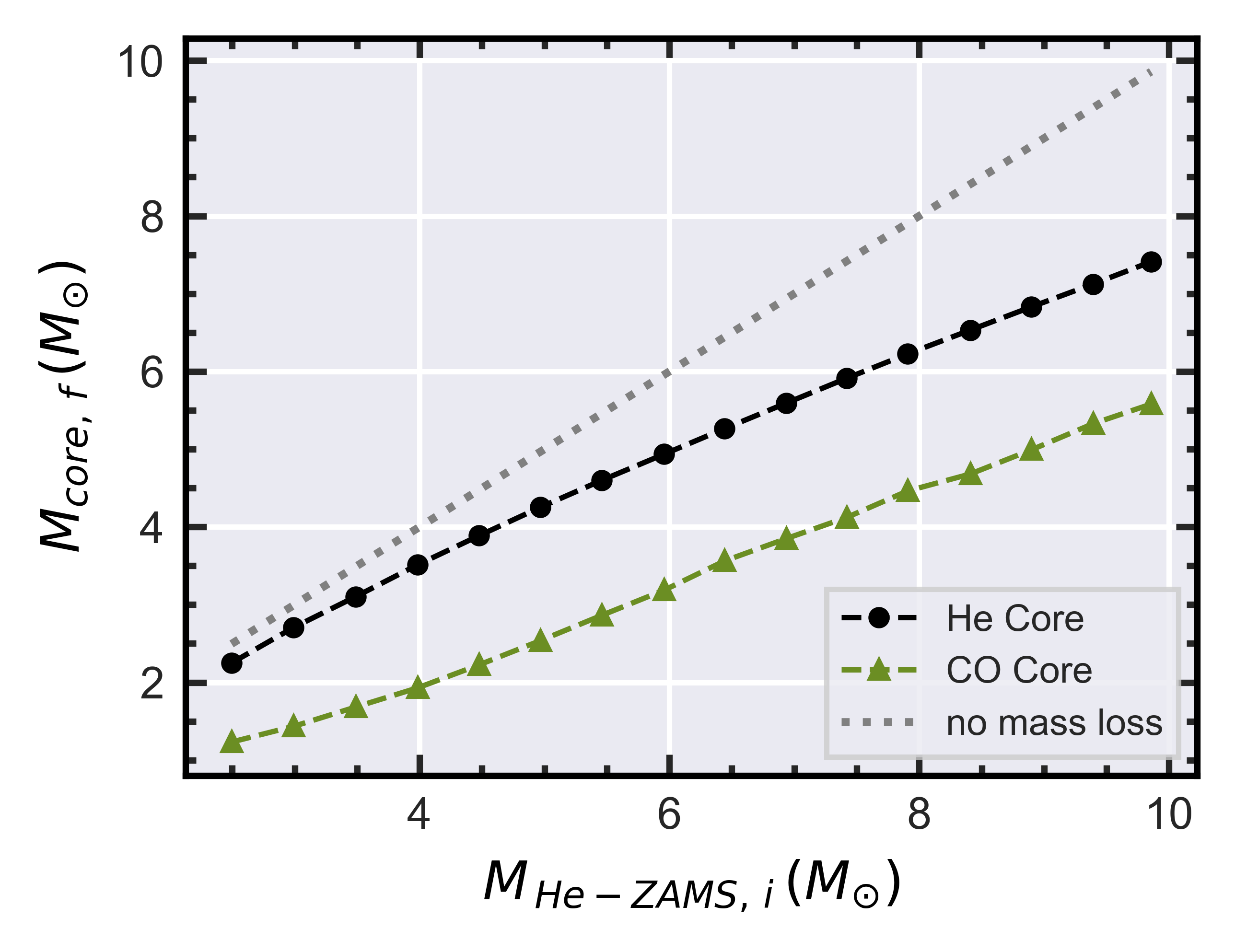}
    \caption{Final helium star core masses as a function of ZAMS mass, for single helium stars with masses in the range $2.5$--$9.8$\,\msun{}. 
    The black dots represent the final He core mass, and the green triangles represent the final CO core mass. The dotted black line shows the line where the two quantities are equal, corresponding to no mass loss. }
    \label{fig:co_he_core_mass}
\end{figure}

Fig~\ref{fig:co_he_core_mass} shows the final He and carbon-oxygen (CO) core masses as a function of initial He-ZAMS mass for all pre-SN stars. There is a linear relation between CO core mass and initial He-ZAMS masses. A similar trend is found in \citet{2018ApJ...860...93S}. The final CO cores are well determined by the initial mass of the helium star and the choice of stellar physical processes. 

In close binary systems, Case BB MT marks the final episode of mass transfer and plays a key role in the binary's evolution for two key reasons. First, it strips the residual helium envelope from the helium star donor, leaving behind an almost naked CO core prior to its SN explosion. Second, depending on the initial mass of the helium star, accretion onto the first-born neutron star during this phase can lead to its recycling. During Case BB MT, Roche lobe overflow governs both the orbital evolution and the mass-stripping process of the helium star.

\subsection{Initial conditions for the grid of He-NS binaries}
\label{subsec:initial_grid}

\begin{figure}
    \centering
    \includegraphics[]{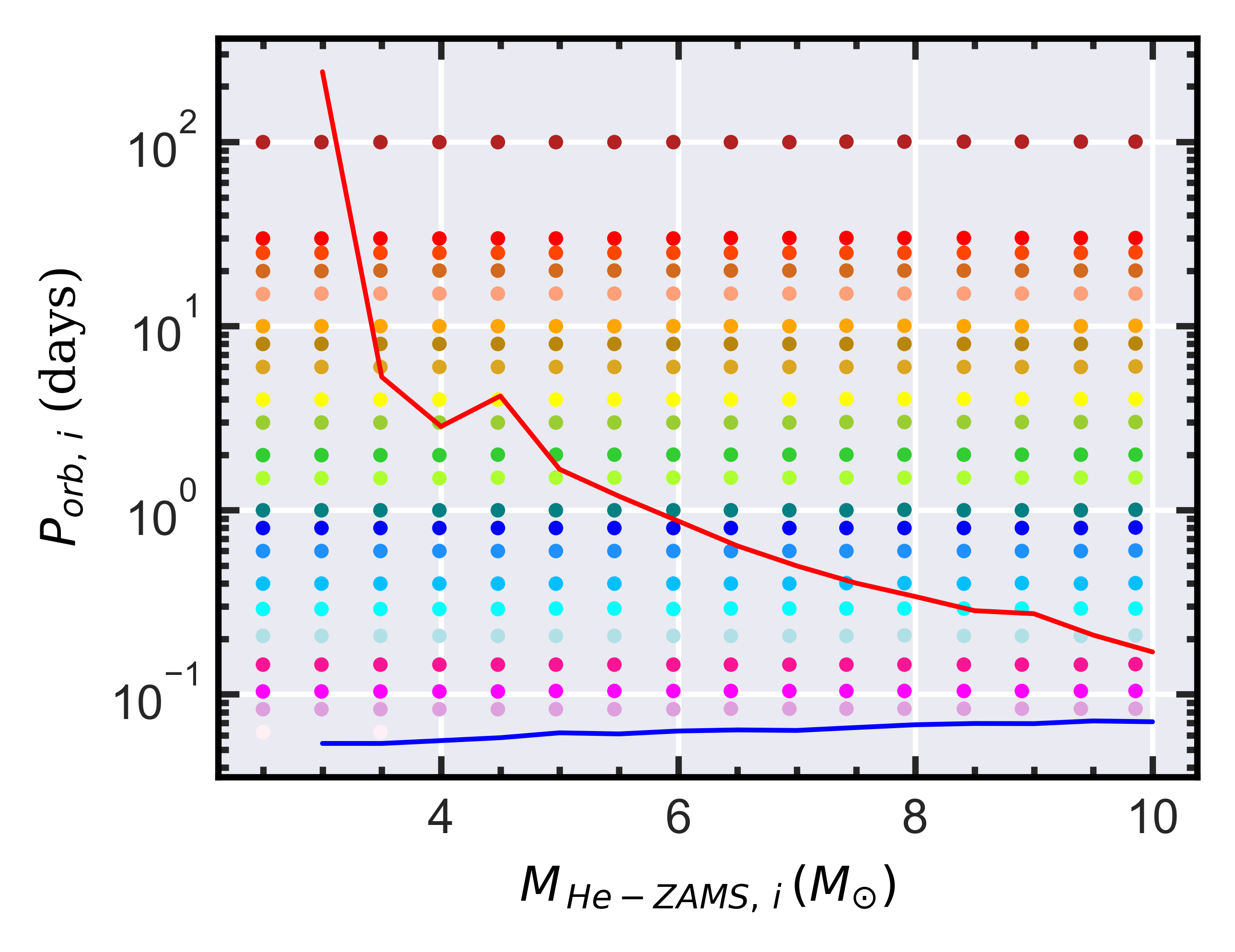}
    \caption{Grid of initial orbital period of He-NS binaries as a function of initial He-ZAMS masses. The red and blue lines show the range of orbital periods that allow mass transfer by Roche lobe overflow. The different colors of the solid circles represent different initial orbital periods.}
    \label{fig:Grid_of_Porb-he_mass}
\end{figure}

For a detailed population study of DNS mergers, a grid of over 338 detailed models of helium star-neutron star binaries was simulated based on the initial masses and orbital period. The initial He mass ($M_\mathrm{{He-ZAMS,\,i}}$) ranges from $2.5$--$9.8$\,\msun{}, with $\Delta M_\mathrm{{He\,i}} = 0.5$. Although the He-star models were initialized with masses ranging from $2.5$--$10$\,\msun{}, they were evolved from pre-main sequence (pre-MS) to the He-ZAMS, during which mass loss via stellar winds was active. As a result, the He-star models ($> 3$\,\msun{}) undergo slight mass loss before reaching He-ZAMS, and thus the final ZAMS He-star masses span $2.5$--$9.8$\,\msun{}. All of our He-stars are evolved at solar metallicity, and all binaries have $M_\mathrm{NS} = 1.4$\,\msun{}.
The orbital period range was calculated based on the parameter space we obtained from the evolution of the He-stars from zero-age main sequence until they expand to their maximum radius (see Fig~\ref{fig:rad_as_fun_of_ini_mass}).
The minimum and maximum range of orbital periods for binaries that will interact and undergo mass transfer is calculated based on the parameter space defined by the maximum and minimum radii of the He-star. For this relation, it is important to understand the relationship between the star's radius and its Roche lobe radius in a binary system. The Roche lobe radius depends on the orbital separation, which is, in turn, related to the orbital period via Kepler's third law. Following this method, we use the fit to Roche lobe radius given by \citet{Eggleton:1983ApJ}. 
Fig \ref{fig:Grid_of_Porb-he_mass} shows a grid plot of the initial orbital period ($P_{\mathrm{orb, i}}$) as a function of initial He-ZAMS mass $M_\mathrm{{He-ZAMS,\,i}}$ with orbital periods ranging from $\sim 0.06$ to $100$ days(d) to have a representative combination of close, semi-detached, and wide orbits.

\subsection{Supernova and kick prescription}
\label{subsec: SN_kick_prescription}

\begin{figure*}
\subfigure[]
{\includegraphics[]{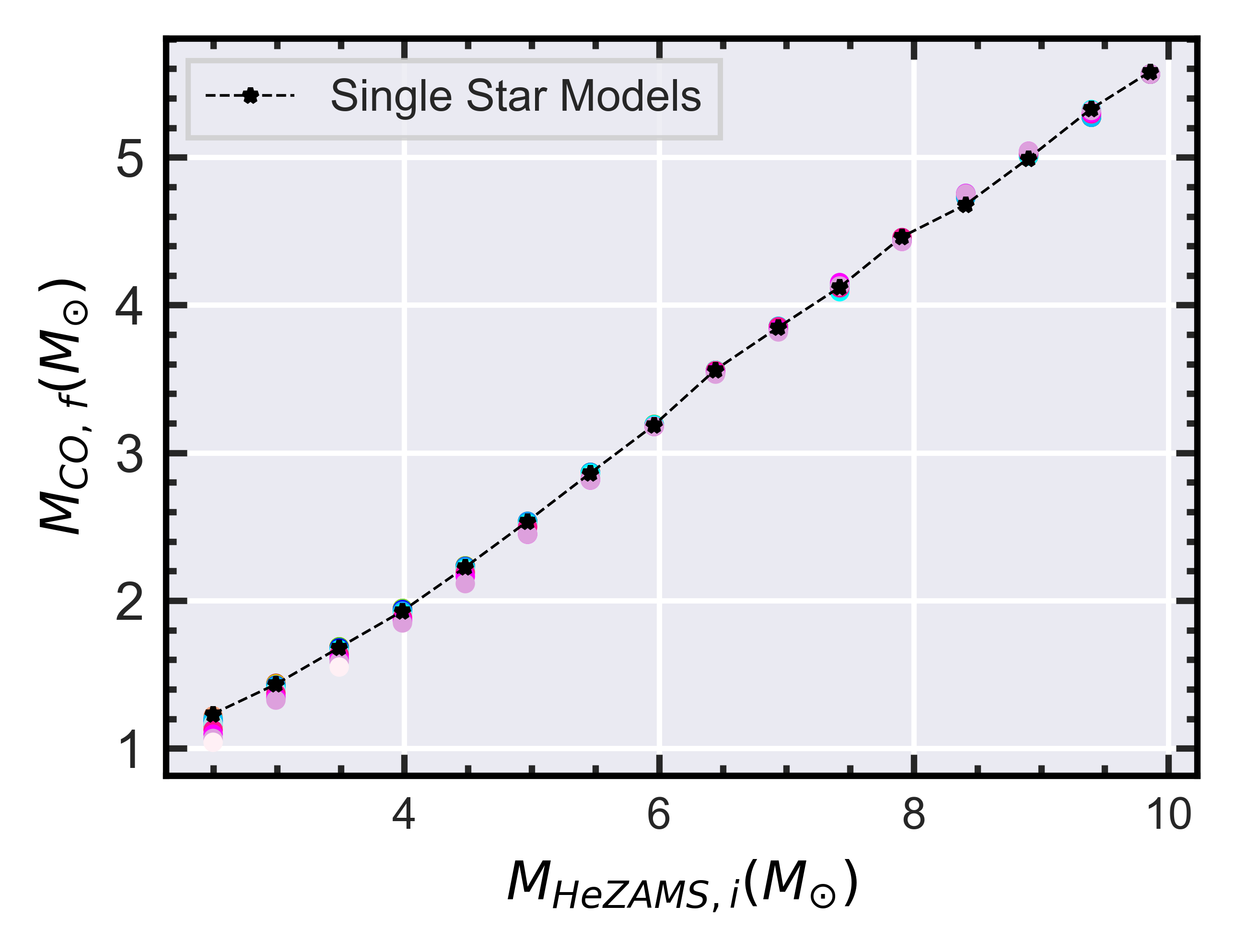}\label{fig:co_core_masses_binary}}
\hspace{0.5cm}
\subfigure[]
{\includegraphics[]{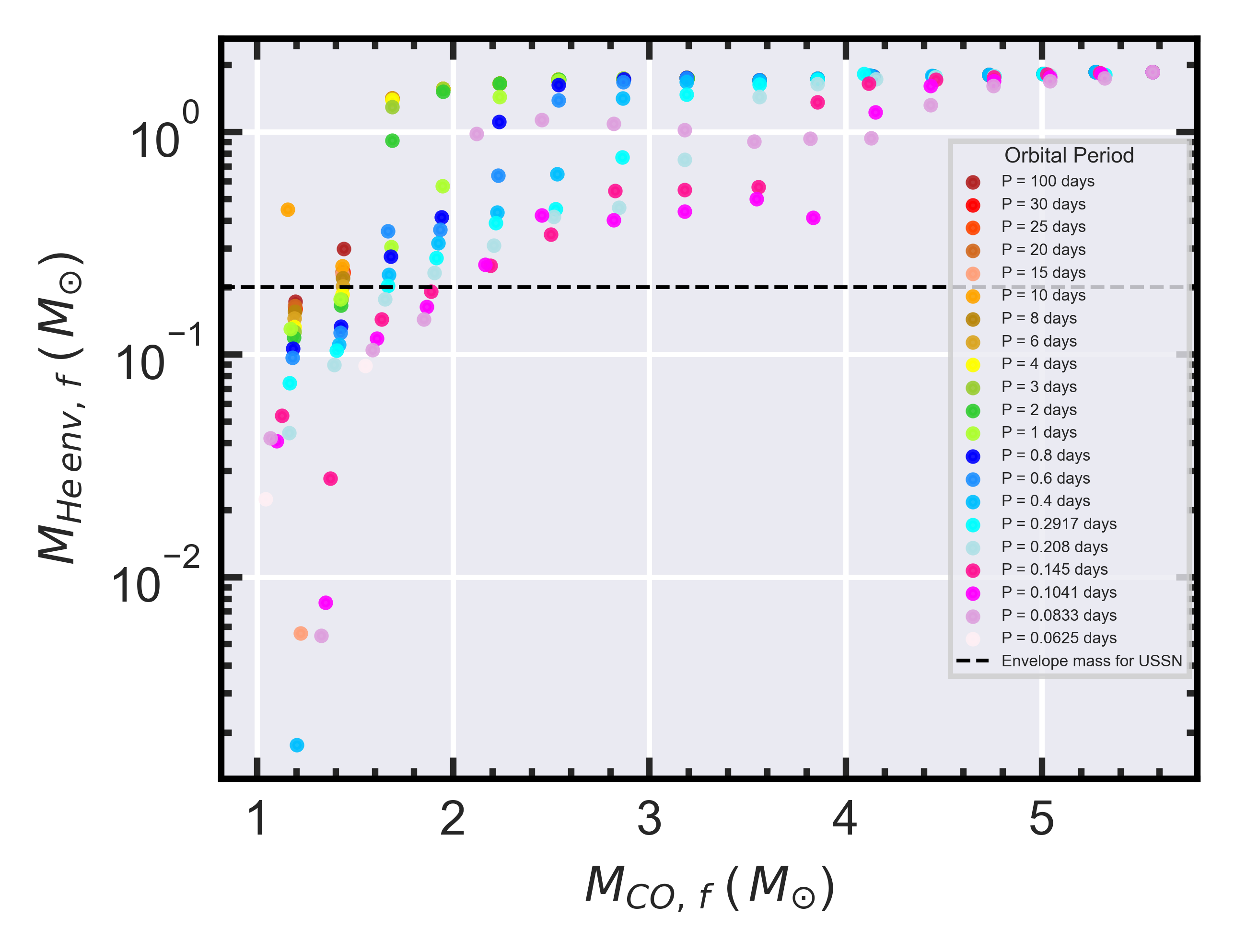}\label{fig:final_helium_env_mass}}
\hspace{0.5cm}
\caption{(Left) Carbon-oxygen (CO) metal core as a function of initial He-ZAMS mass. The black points are the CO core masses of single-star models (as shown in Figure~\ref{fig:co_he_core_mass}), whereas the different colored points are CO core masses from binary grid models. (Right) The final helium envelope mass as a function of the final CO core mass of pre-SN models is shown in the right panel. The black dashed line highlights the helium envelope mass condition to define ultra-stripped SNe as exploding stars in close binaries defined as $M_\mathrm{He,env,f} \leq  0.2$\,\msun{} in \citet{Tauris:2015MNRAS} 
}
 \label{fig:final_co_core_mass_and_he_envelope}
\end{figure*}

After the He star has transferred mass, it leaves behind a pre-SN carbon-oxygen (CO) metal core (see fig~\ref{fig:co_core_masses_binary}) and some helium envelope (see fig~\ref{fig:final_helium_env_mass}). Following the final stages of nuclear burning, the core is expected to collapse and form a compact object. 
In Section~\ref{subsec:He_star_in_DNS}, we discussed how a low-mass helium star expands to a greater dimension and undergoes a prolonged phase of mass transfer. 
Especially when the binary is in close orbit (with $P_\mathrm{orb,i} \leq 1$\,day), the mass transfer can last longer and leave behind an almost naked CO-core; naked here mean only a little or negligible helium envelope. 
This kind of system with low-mass helium stars explodes as an ultra-stripped supernova (USSN). \citet{Tauris:2015MNRAS} defines USSN as an explosion of a He star in a close binary orbit with a final helium envelope of $\leq\,0.2$\,\msun{}. The ultimate fate of the close binaries depends on the mass of the matter ejected during the SN explosion and the kick velocity received by the newborn neutron star. This altogether is dependent on the type of SN explosion. The USSN can also be an electron-capture SN (ECSN) or iron core collapse SN (Fe CCSN). 
Only a narrow range of He-ZAMS stars, with the final ONe core mass in the range of 1.37 - 1.43\,\msun{} \citep[][]{1984ApJ...277..791N,2004ApJ...612.1044P,Takahashi_2013,Tauris:2015MNRAS} are thought to undergo ECSN. 
As these explosions have low energies and the progenitor star has a low helium envelope mass, this leads to the newborn NS receiving a low kick of only a few tens of $\mathrm{kms^{-1}}$ \citep{2004ApJ...612.1044P,2006A&A...450..345K,Tauris:2015MNRAS,2006ApJ...644.1063D,Gessner:2018ApJ,Guo:2024MNRAS}. 
Whereas if the final ONeMg metal core exceeds $\sim 1.43$\,\msun{}, during the carbon burning, the core will ignite oxygen and lead to a usual Fe CCSN, which has stronger explosion energies and kick the newborn neutron stars up to a few thousands of $\mathrm{kms^{-1}}$ \citep{2012ARNPS..62..407J, 2013A&A...552A.126W}. 

To calculate the remnant masses and kick velocity post-SN explosion, we use an analytical prescription given by 
\citet[][]{Mandel:2020MNRAS}. We use the calibration given by \citet{Kapil:2023MNRAS}. The prescription calculates the dependence of the remnant mass of NS and BHs and natal kicks based on the pre-SN carbon-oxygen (CO) core mass and helium envelope mass. This recipe accounts for the intrinsic stochasticity of stellar evolution and the SNe mechanism by providing a probabilistic prescription. A detailed explanation can be found in \citet{Mandel:2020MNRAS}. 

\begin{figure*}
\subfigure[Neutron star remnant masses]
{\includegraphics[]{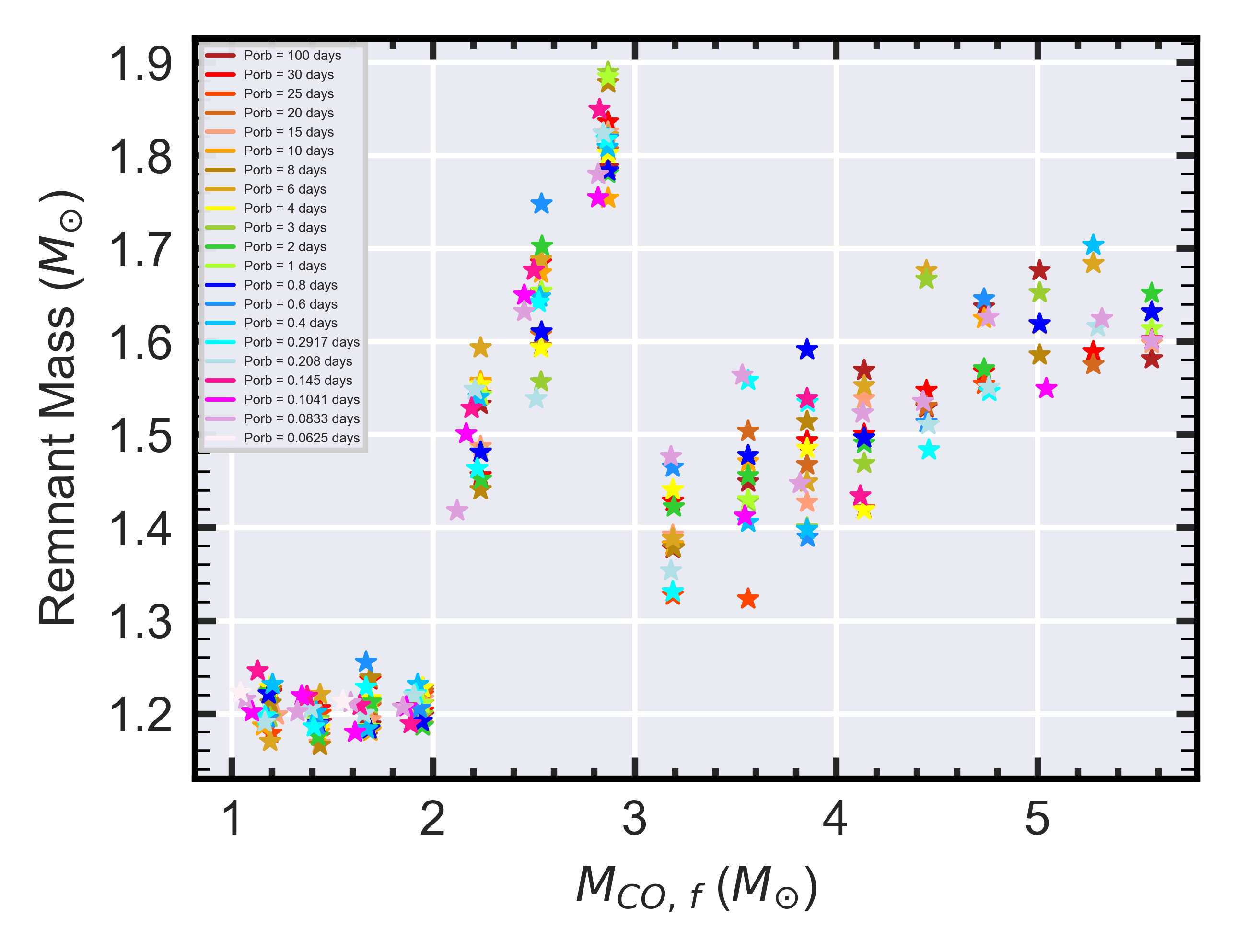}\label{fig:remnant_masses_std_model}}
\hspace{0.5cm}
\subfigure[Neutron star natal kick velocities]
{\includegraphics[]{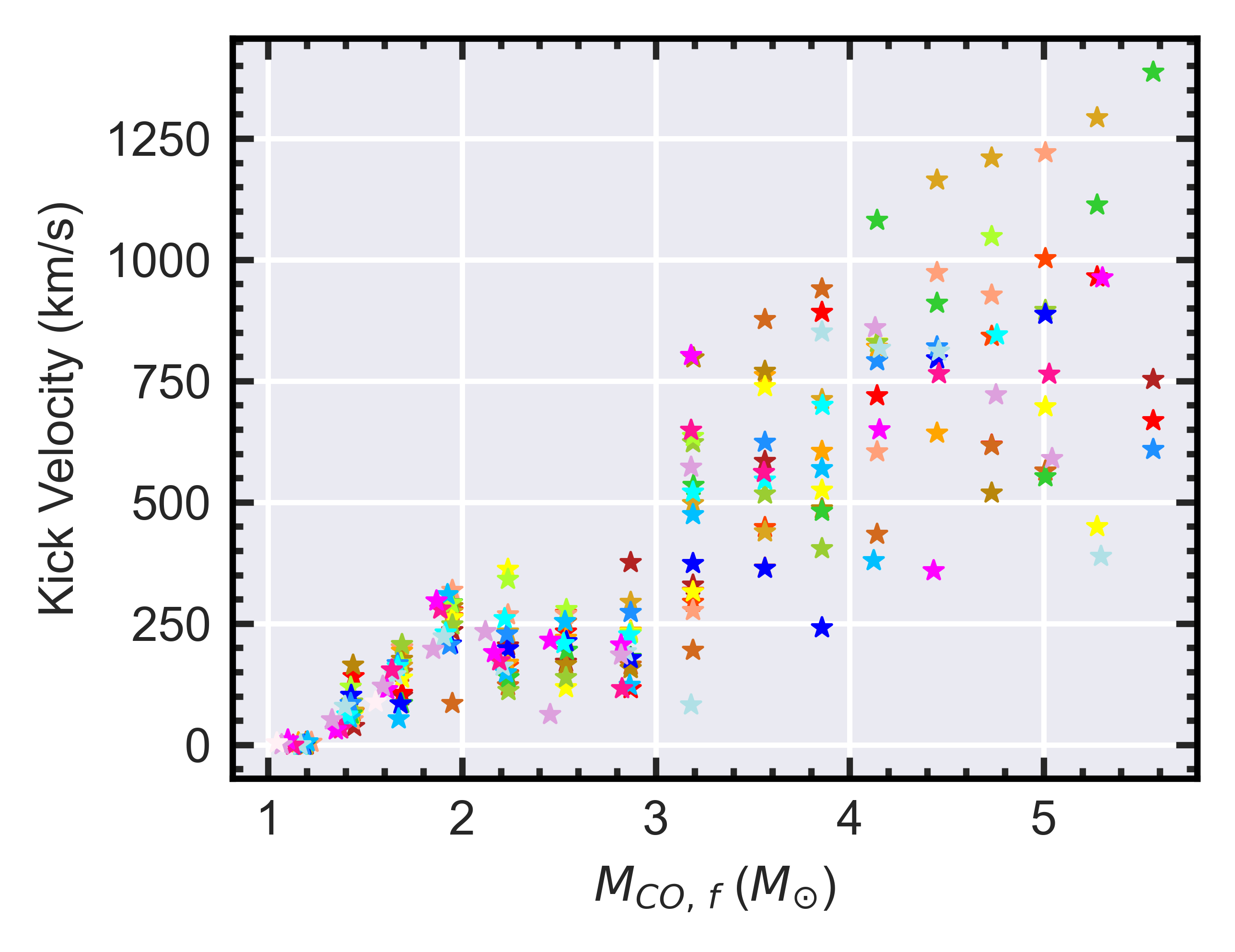}\label{fig:std_model_kick_velocities}}
\hspace{0.5cm}

 \caption{(Left) Neutron star remnant masses and (right) natal kick velocity in a He-NS binary as a function of CO core masses calculated using the standard model \citep[][]{Mandel:2020MNRAS}. The different coloured stars represent different initial orbital periods.}
 
 \label{fig:standard_remnant_mass_kick_vel}
\end{figure*}

The NS kicks in the \citet{Mandel:2020MNRAS} model can be parameterized by two parameters: $v_\mathrm{{ns}}$, which is a scaling prefactor for NS kicks that defines the relation between the final CO core mass and the new born remnant mass, and $\sigma_{\mathrm{kick}}$, a measure of the scatter in the kick distribution. \citet{Kapil:2023MNRAS} uses observational data of single pulsar velocities to constrain these parameters for NS natal kicks and study some properties of the resulting kick distribution. They also study the inference of those parameters from binary stellar evolution models. \citet{Mandel:2020MNRAS} model also applies the NS natal kicks to ECSNe and USSNe. Based on this detailed study, we use one of the best-fitting parameters - $v_{\mathrm{NS}} = 525$\,km s$^{-1}$ and $\sigma_{\mathrm{NS\, kick}} = 0.3$. We refer to this prescription as a \textit{`Standard Model'}.
Fig~\ref{fig:standard_remnant_mass_kick_vel} shows the corresponding remnant masses and natal kick velocity as a function of pre-SN carbon-oxygen core.

\subsection{Post-SN orbital dynamics}
\label{Post-SN_orbital_dynamics}

The pre-SN binary with CO metal core and the firstborn NS, with pre-SN orbital separation $a_{i}$  and orbital period $P_{\mathrm{orb,\,pre-SN}}$, undergoes an instantaneous SN explosion \citep{Brandt1994TheEO}. The eccentricity and orbital period of the binaries change due to sudden mass loss and kick velocity imparted onto the post-SN remnant due to explosion asymmetry, without affecting the separation, because SNe in binary systems occur on a timescale much shorter than the orbital period. Fig 1 from \citet{Brandt1994TheEO} shows the binary system and supernova kick geometry. The remnant NS is kicked in a direction specified by two angles, $\theta$ and $\phi$, where $\theta$ is the angle between the direction of the kick velocity vector and the initial orbital plane, and $\phi$ is the angle between the initial direction of motion of the CO core companion and the projection of the kick velocity vector on the orbital plane. 
We assume that the kick direction is randomly directed, drawing $\theta$ and $\phi$ from an isotropic distribution.
For the system to remain bound post-SN, the post-SN energy has to be negative.
We use Eq. (2.1) - (2.8) from \citet{Brandt1994TheEO} to calculate the post-SN separation, orbital period, and eccentricity of the DNS binaries.

\subsection{Gravitational wave merger timescale}
\label{subsec:gw_merger_timescale}

Now that we have gravitationally bound DNS systems, we calculate how long it would take for them to merge due to the emission of gravitational waves.
\citet{Peters:1964PhRv} derived a post-Newtonian expression for the time $T$ for two point masses to spiral in through gravitational wave emission from an initial separation $a_{0}$ and initial eccentricity $e_{0}$. 
To calculate the same, we use the analytic fit from \citet{Mandel:2021RNAAS} given to the duration of a two-point mass driven by gravitational wave emission. 
\begin{equation}
    T \approx T_{c}(1+0.27 e^{10}_{0}+0.33 e^{20}_{0} + 0.2 e^{1000}_{0})(1-e^2_{0})^{7/2}
\end{equation}
 where $T_\mathrm{{c}}$ is defined as the finite time in which the system decays for $e = 0$ ( \citep[see eq. 5.10 in][]{Peters:1964PhRv} 

\subsection{Pulsar recycling}
\label{subsec:recycling}

We assume that only pulsars that have experienced mass transfer and have been recycled will be observed as radio pulsars due to the longer lifetimes of these systems compared to non-recycled pulsars \citep[e.g.,][]{Chattopadhyay:2020MNRAS}. However, non-recycled DNSs can still contribute to the gravitational wave population. 

\section{Results}
\label{sec:Results}

\subsection{Population synthesis}
\label{sec:population_synthesis}
\begin{figure*}
\subfigure[\mhe{} = 2.9 \msun{} ]
{\includegraphics[]{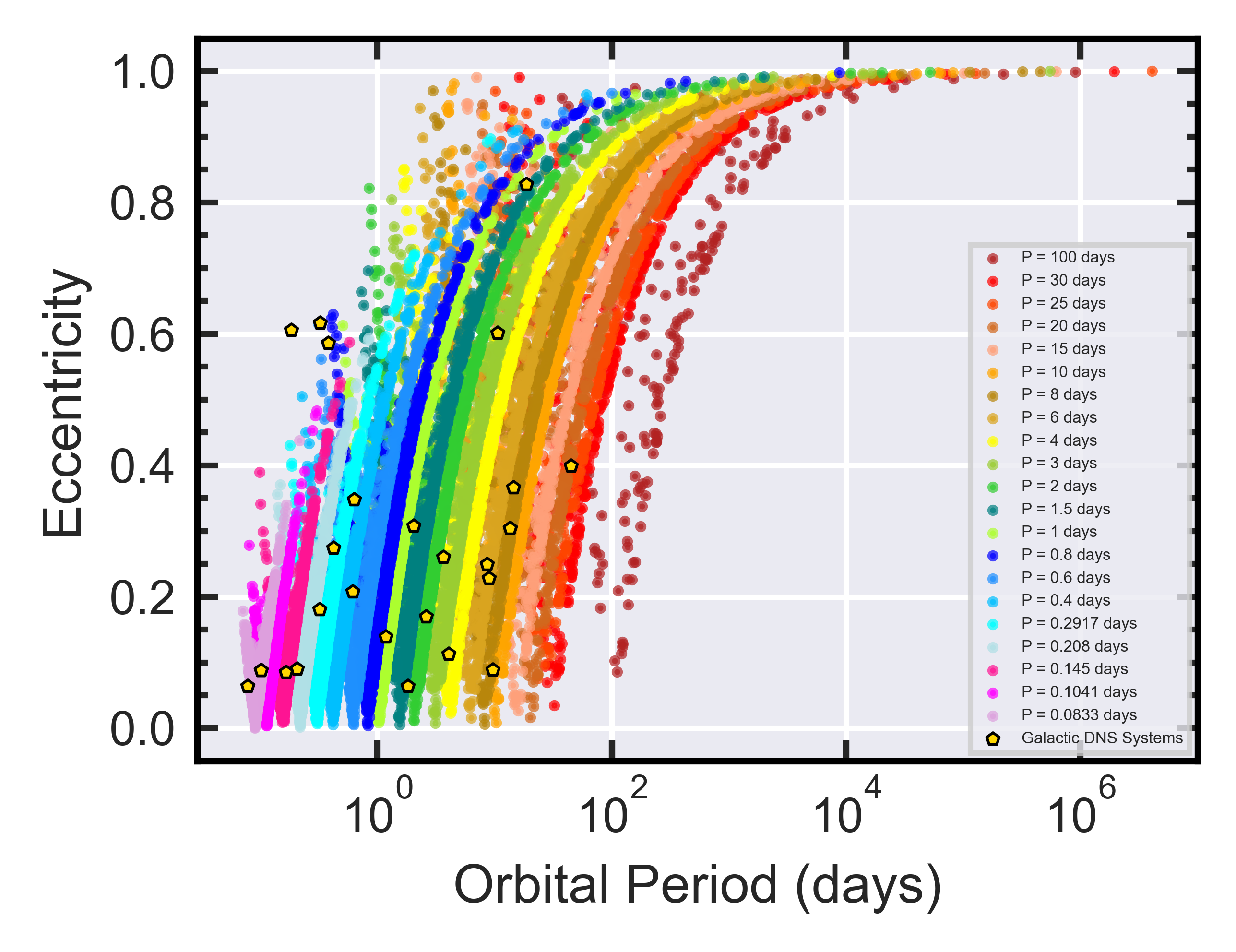}\label{fig:Porb-e_for_2.9M}}
\hspace{0.5cm}
\subfigure[\mhe{} = 3.9 \msun{} ]
{\includegraphics[]{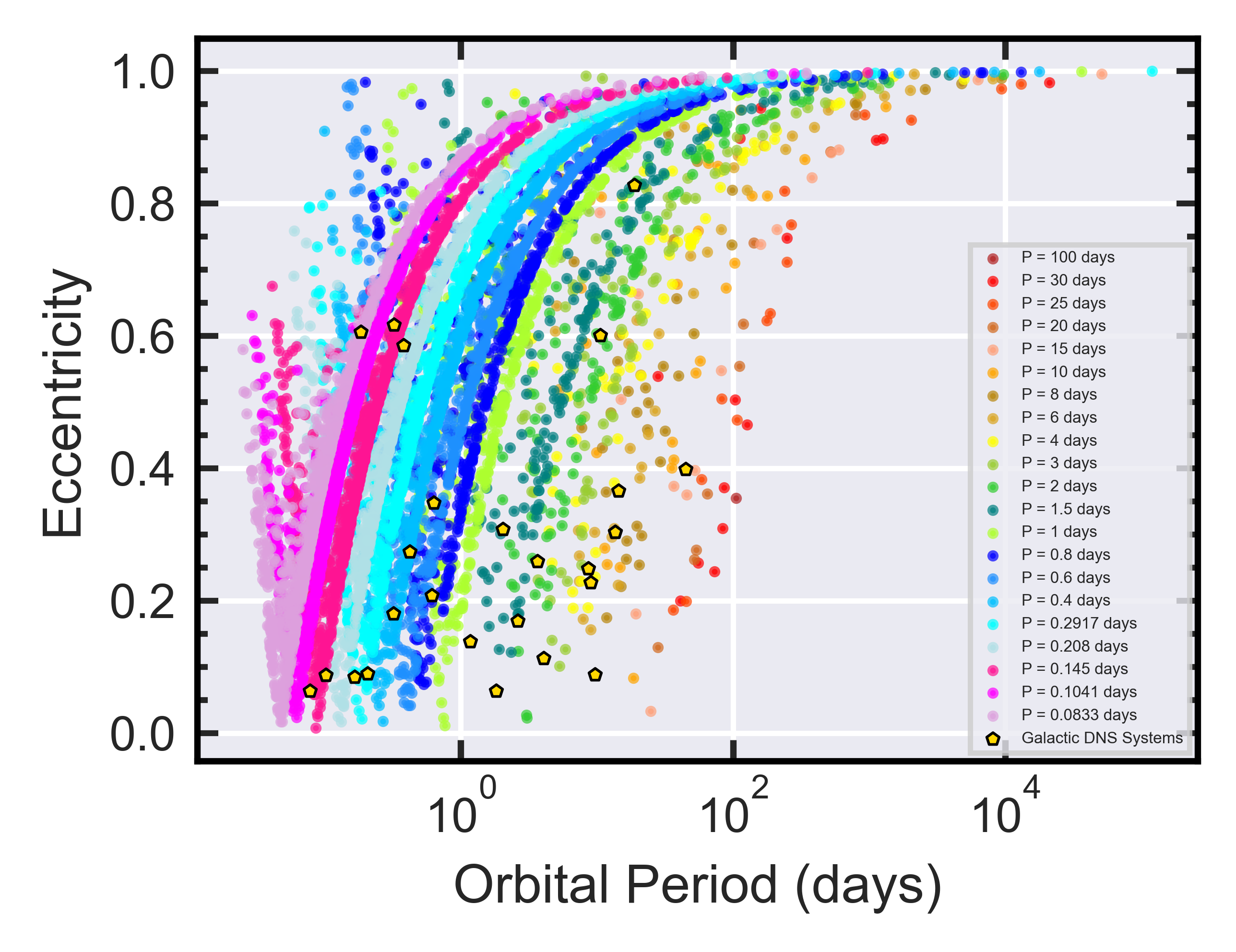}\label{fig:Porb-e_for_4.0M}}
\hspace{0.5cm}
\subfigure[\mhe{} = 5.4 \msun{} ]
{\includegraphics[]{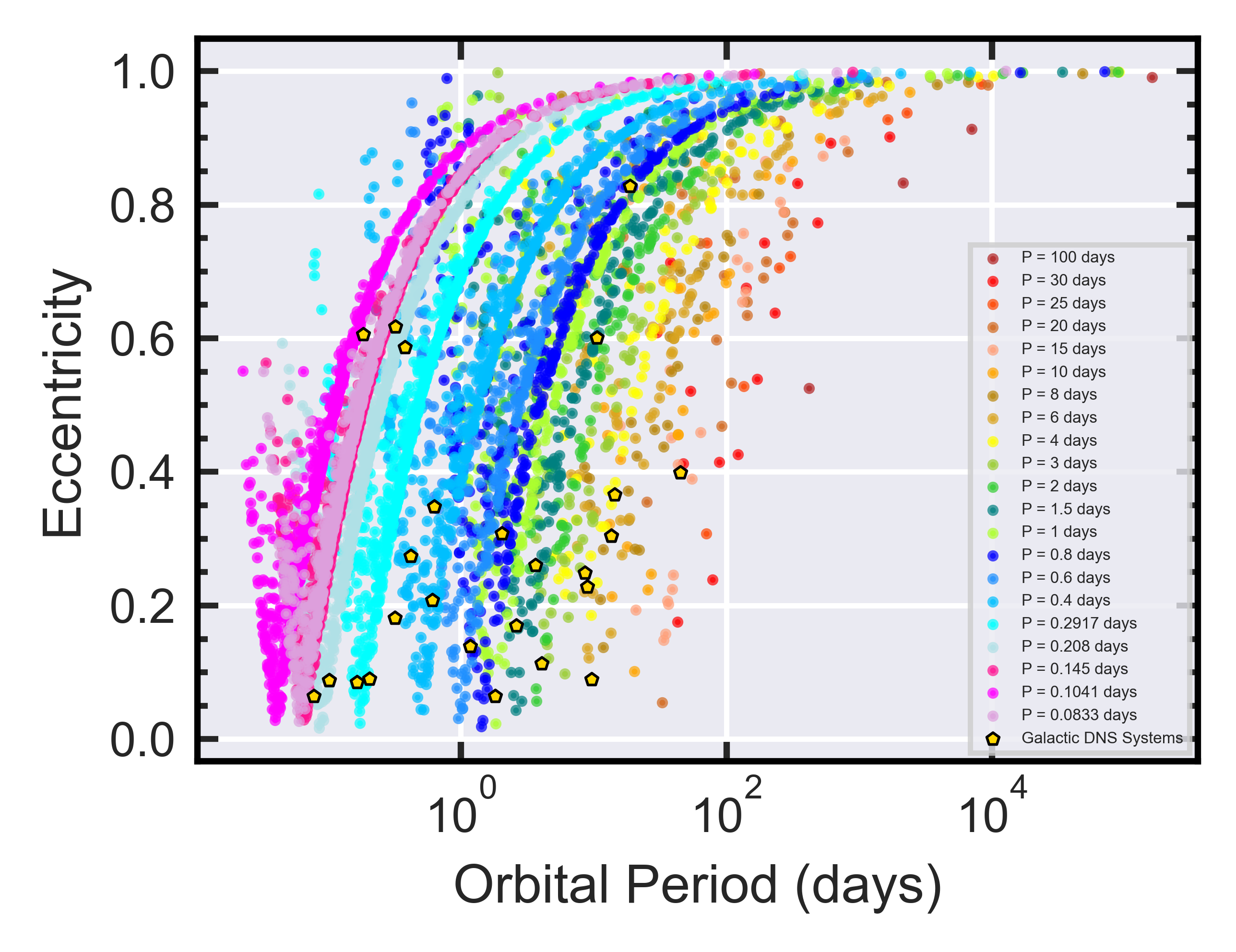}\label{fig:Porb-e_for_5.5M}}
\hspace{0.5cm}
\subfigure[\mhe{} = 8.9 \msun{} ]
{\includegraphics[]{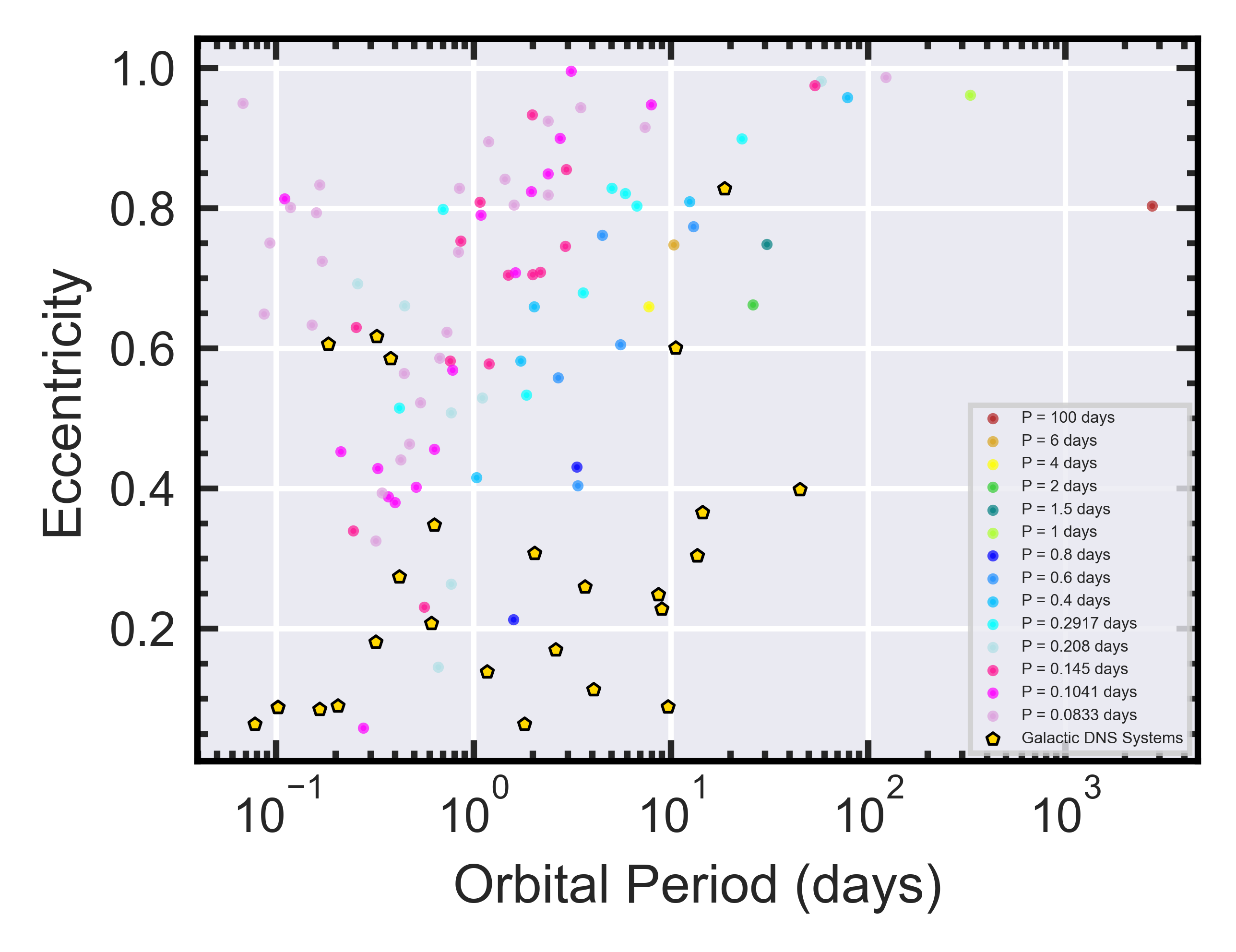}\label{fig:Porb-e_for_8.9M}}
\hspace{0.5cm}

 \caption{Distribution of the simulated post-SN systems in the $P_{\mathrm{orb}}-e$ diagram for helium stars of masses 2.9\,\msun{}, 3.9\,\msun{}, 5.4\,\msun{}, 8.9\,\msun{}, using the standard model of SN prescription, in which we performed 1000 cycles with random isotropically oriented NS kicks for each pre-SN system. 
 The different colors correspond to the initial orbital periods ranging from 0.06--100\,d. The yellow stars represent the confirmed Galactic DNS systems in the field.}
 \label{fig:std_porb-e_of_binaries}
\end{figure*}

\subsubsection{Standard Model}
\label{sec:standard_model}

\begin{center}
    \textbf{A. $P_\mathrm{orb}$-e distribution}\\
\end{center}

\begin{figure}
    \centering
    \includegraphics[]{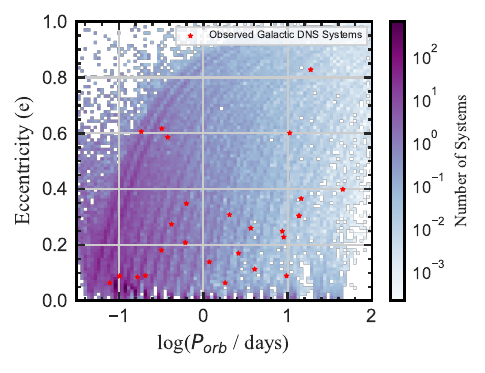}
    \caption{Orbital period–eccentricity distribution ($P_\mathrm{orb}$-e) of simulated post-supernova DNS systems. The 2D histogram shows the predicted distribution based on the standard supernova (SN) prescription model from \citet{Mandel:2020MNRAS} and \citet{Kapil:2023MNRAS}. The color bar on the right-hand side represents the number of systems increasing from light blue to purple. We simulate 338 He-NS binaries with $P_\mathrm{orb,\,i}$ that vary logarithmically from $0.06$--$100 $\,d. The red stars are the observed Galactic DNS systems.}  
    \label{fig:std_model_2d_porb-e}
\end{figure}

Fig~\ref{fig:std_porb-e_of_binaries} shows the final distribution of eccentricity as a function of the final orbital period produced due to the dynamical effects of applying a kick to our stellar models. We perform 1000 trials with randomly oriented isotropic kicks. 

Fig~\ref{fig:Porb-e_for_2.9M} shows the $P_\mathrm{orb}$-e distribution for a binary with a $1.4$\,\msun{} NS and $2.9$\,\msun{} He star companion, along with the 24 confirmed observed Galactic DNS systems in the field. After the mass transfer phase, the He star leaves behind a CO metal core of masses $1.33$\,\msun{}--$1.44$\,\msun{} depending on the initial orbital period. The SN explosion leaves behind a remnant of mass $\sim 1.3$\,\msun{}. This is an example of a low-mass He star in a He-NS binary. 
Such systems with low-mass helium stars explode as USSNe and ECSNe in close orbits. Low-mass helium envelopes lead to fast explosions and low explosion energies, resulting in kick velocities of only a few tens of $\mathrm{kms^{-1}}$. Close binaries, in turn, yield small post-SN eccentricities. In wider systems, less mass is transferred prior to core collapse, leaving behind a slightly more massive helium envelope ($\geq$\,$0.2$\,\msun{}) at the time of explosion, typically resulting in a Fe CCSN (see fig~\ref{fig:final_helium_env_mass}). The larger envelope mass ejected during the supernova leads to stronger kicks imparted onto the newborn neutron star, resulting in higher velocities and a wide range of post-SN eccentricities (from $0$--$1$). Many wide binaries are either disrupted or kicked into highly eccentric orbits.

Fig~\ref{fig:Porb-e_for_4.0M} and \ref{fig:Porb-e_for_5.5M} shows the $P_\mathrm{orb}$-e plot for more massive binaries, consisting of a NS with an intermediate-mass He star companion of $3.9$\,\msun{} and $5.4$\,\msun{}, respectively. 
Intermediate and massive helium (fig.~\ref{fig:Porb-e_for_8.9M}) stars have shorter lifetimes, which lead to the expansion of the stars on a shorter timescale, resulting in shorter mass transfer phases and the retention of relatively massive helium envelopes ($\sim 0.2$--$1$\,\msun{}). Consequently, the resulting neutron stars receive large natal kicks, typically in the range of $500$--$1500$\,$\mathrm{kms^{-1}}$. For each helium star mass, we compute the fraction of post-supernova systems that remain gravitationally bound by sampling a wide range of initial orbital periods ($0.06$\,d--$100$\,d). For example, in the case of a $3.9$\,\msun{} model, approximately $\sim 34$\,\% of the systems survive, thereby majorly disrupting wide binaries with $P_\mathrm{orb} > 1$\,d. Close binaries receive higher kick velocities, producing a broad distribution of post-supernova eccentricities. For the most massive He stars considered (fig.~\ref{fig:Porb-e_for_8.9M}), only $1$\,\% systems remain gravitationally bound, the majority of binaries are disrupted by the large natal kicks. We note that in a significant fraction of the disrupted systems, the second-born compact object is a black hole.

Fig~\ref{fig:std_model_2d_porb-e} represents the combined distribution of the simulated post-SN DNS systems in the $P_\mathrm{orb}$-e plane, and the observed Galactic DNS systems represented as red stars. 
We assume a simple model for the initial distributions of He star-NS binaries. For the initial orbital period, we assume a ``flat in the log" distribution $(P(P_\mathrm{orb}) \sim 1/P_\mathrm{orb})$. For the distribution of initial He star masses we assume a ``power law" distribution $(p(M_\mathrm{He}) \sim M_\mathrm{He}^{-\alpha}$, inspired by the stellar initial mass function \citep[IMF;][]{1955ApJ...121..161S}, where $\alpha = 2.35$.
The number of DNS systems per bin in the $P_\mathrm{orb}$--e plane is calculated using a 2D histogram weighted by the assumed initial distribution of He–NS binaries. Each simulated binary is assigned a statistical weight based on its initial parameters — the initial mass of the helium star and the initial orbital period. These weights are derived using the initial distribution model described above. The 2D histograms are then normalized using these weights, and the resulting color scale reflects the relative number of systems in each bin. We find that the trends in the $P_\mathrm{orb}$--e plane are robust across a range of assumptions. In future work, we plan to interface this model with a full rapid binary population synthesis code like COMPAS \citep{Stevenson_2017,Riley_2022} or COSMIC \citep{2020ApJ...898...71B} to derive a more self-consistent initial distribution.

The observed distribution of DNS systems in the field in the $P_\mathrm{orb}$–$e$ plane shows a noticeable gap in eccentricities between $e = 0.3$--$0.6$, which is not reproduced by our standard model. Using the standard SN prescription, our simulations predict that even very close binaries, with initial periods of just a few hours, can result in a wide range of post-SN eccentricities spanning $e = 0$--$1$. 
In particular, as discussed before, low-mass helium stars ($2.5$--$2.9$\,\msun{}) tend to explode as USSN or ECSN, due to their low ejecta mass, low explosion and rapid explosion timescales.
These close systems receive larger kick velocities $ \sim 50$--$150$\,$\mathrm{kms^{-1}}$ and overpredict eccentricities, particularly for the systems resulting from $3.0$\,\msun{} He stars. 
This arises because the \citet{Mandel:2020MNRAS} model overestimates the kick velocities of ECSNe \citep[][]{Gessner:2018ApJ,Guo:2024MNRAS} and USSNe \citep[][]{Tauris:2013ApJL,Tauris:2015MNRAS,Suwa2015,Muller:2019MNRAS}. 
Their model assumes the same kick mechanism used for CCSN. 
Intermediate-mass helium star models ($3.9$--$5.4$\,\msun{}) tend to populate the short period ($P_\mathrm{orb} < 2$\,d), high eccentricity ($e > 0.4$) region of the $P_\mathrm{orb}$-e parameter space. Meanwhile, wide post-SN binaries descending from He stars  $ \geq 5.9$\,\msun{} are largely disrupted.

This discrepancy leads to two possible interpretations: either the Galactic DNS population in this parameter space has not yet observed a system with such orbital properties via radio pulsar surveys, or the model overestimates eccentricities, likely due to uncertainties in the kick magnitudes associated with different SN explosions.

\begin{center}
    \textbf{B. DNS Total Mass Distribution}
\end{center}

\begin{figure}
    \centering
    \includegraphics[]{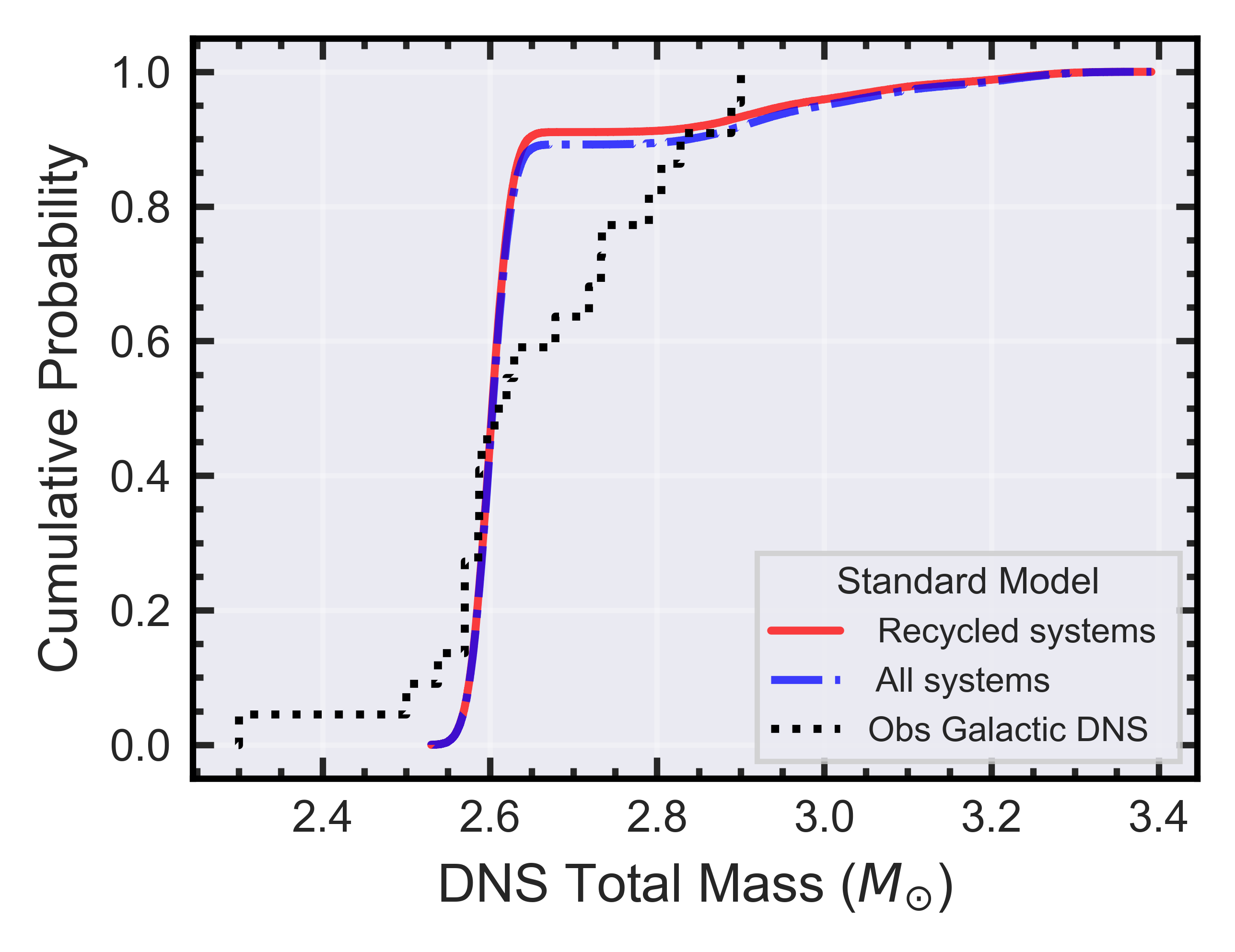}
    \caption{The cumulative distribution of double neutron star total masses. The solid, red line represents the recycled systems based on the standard \citet{Mandel:2020MNRAS} prescription, the blue dot-dashed line represents all formed double neutron star binaries in the simulation based on the standard prescription, and the black dotted line represents the observed Galactic DNS mass distribution.}
    \label{fig:CDF_Mtot_all}
\end{figure}

Our models produce DNS systems with total mass ranging from $2.5$--$3.4$\,\msun{} calculated using the standard prescription, see fig.~\ref{fig:CDF_Mtot_all}).
The red line shows the simulated population of recycled systems, representative of the observed population of DNS systems in the field. 
The blue dot dashed line represents all of the simulated systems, which are compared to the GW systems' population.
Binary systems with lower total masses ($< 2.65$\,\msun) are more common than high masses, due to the assumed weighting scheme. The constant peak in the distribution at $\sim 2.6$\,\msun{} is also influenced by the assumption that the first-born neutron star has a fixed mass of $1.4$\,\msun{}. 
We do not see any DNS systems with total mass between $2.2$--$2.5$\,\msun{} because the smallest CO metal core in our model produces NSs of masses $> 1.2$\,\msun{}, which results in DNSs of total mass $> 2.5$\,\msun{}. 
We observe an extended tail towards high-mass DNSs ($M_{\mathrm{tot}}$) in both the modelled populations; however, we focus on the recycled population, which can be compared to the observed population. However, such an extended tail is not observed in the observed population of DNSs in the field (black dotted line in fig~\ref{fig:CDF_Mtot_all}). 
The high-mass tail in our models comes from the previously mentioned assumption of the CO core masses and remnant masses relation ($\mu_{\mathrm{2b}}$ = $0.5$), in particular the spike that occurs at $M_\mathrm{CO} \sim 2.9$\,\msun{} that produces remnant NS of masses $1.6$\,\msun{} $\leq$ \mns{} $\leq$ $2.0$\,\msun{} which in turn overproduces massive DNS systems. CO core of masses ranging between $2$\,\msun{} $\leq$ \mco{} $<$ $3$\,\msun{} are resultant of intermediate-mass helium stars ($3.9$--$5.9$\,\msun{}). Even the small fraction $1.1$\,\% of systems that survive in $7.9$--$9.8$\,\msun{} He-star mass range tend to produce heavier DNS systems, with total masses exceeding $3.0$\,\msun{}  This can mean two things – either we have not seen any massive systems in radio observations yet, or that there is an indiscrepancy between the modeled and observed population.
A similar trend has been observed in previous population synthesis studies \citep[e.g.,][]{Vigna-Gomez:2018MNRAS, Chattopadhyay:2020MNRAS,Mandel:2021MNRAS}.

\subsubsection{Modified model}

\begin{figure*}
\subfigure[Neutron star remnant masses]
{\includegraphics[]{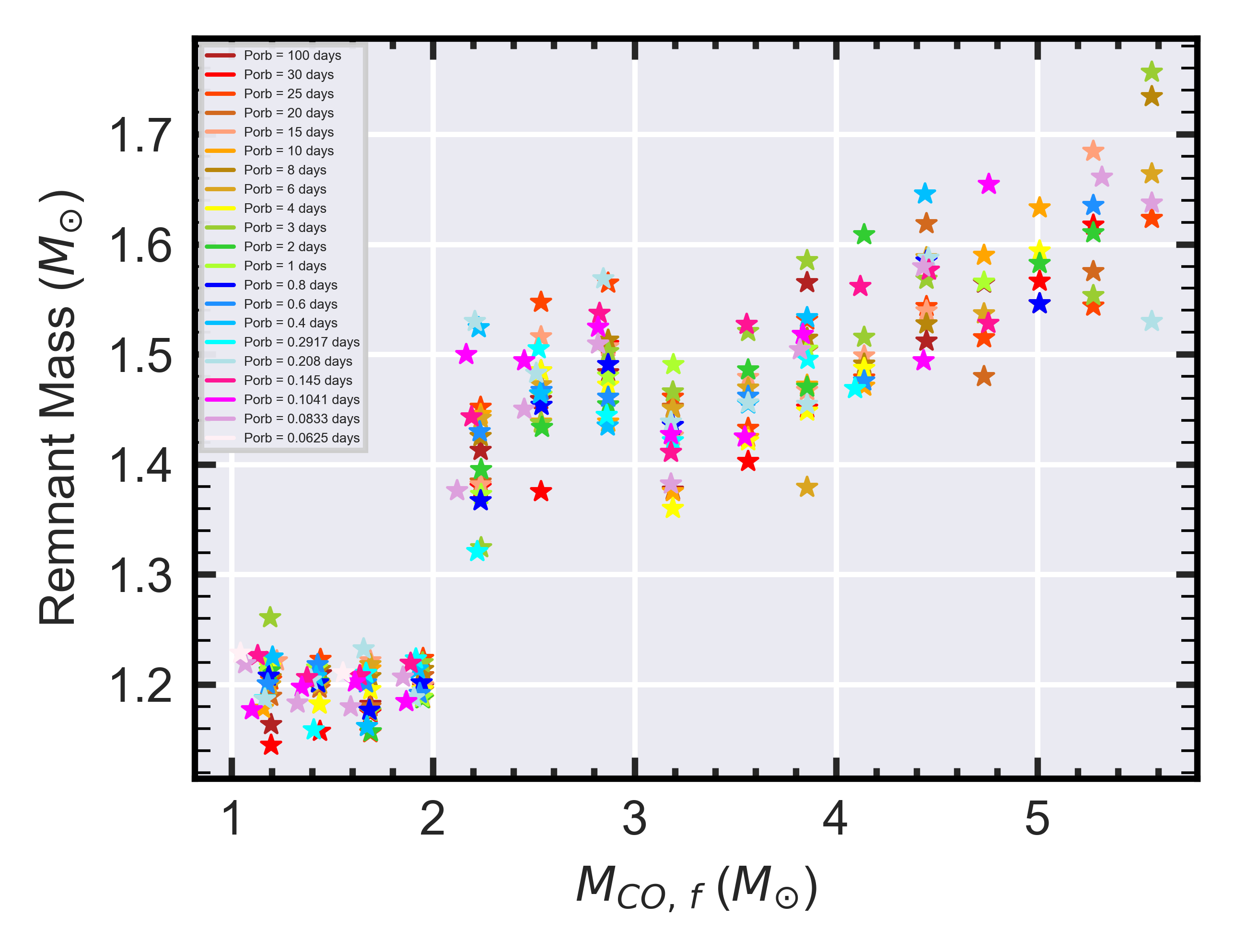}\label{fig:modified_remnant masses}}
\hspace{0.5cm}
\subfigure[Neutron star natal kick velocities]
{\includegraphics[]{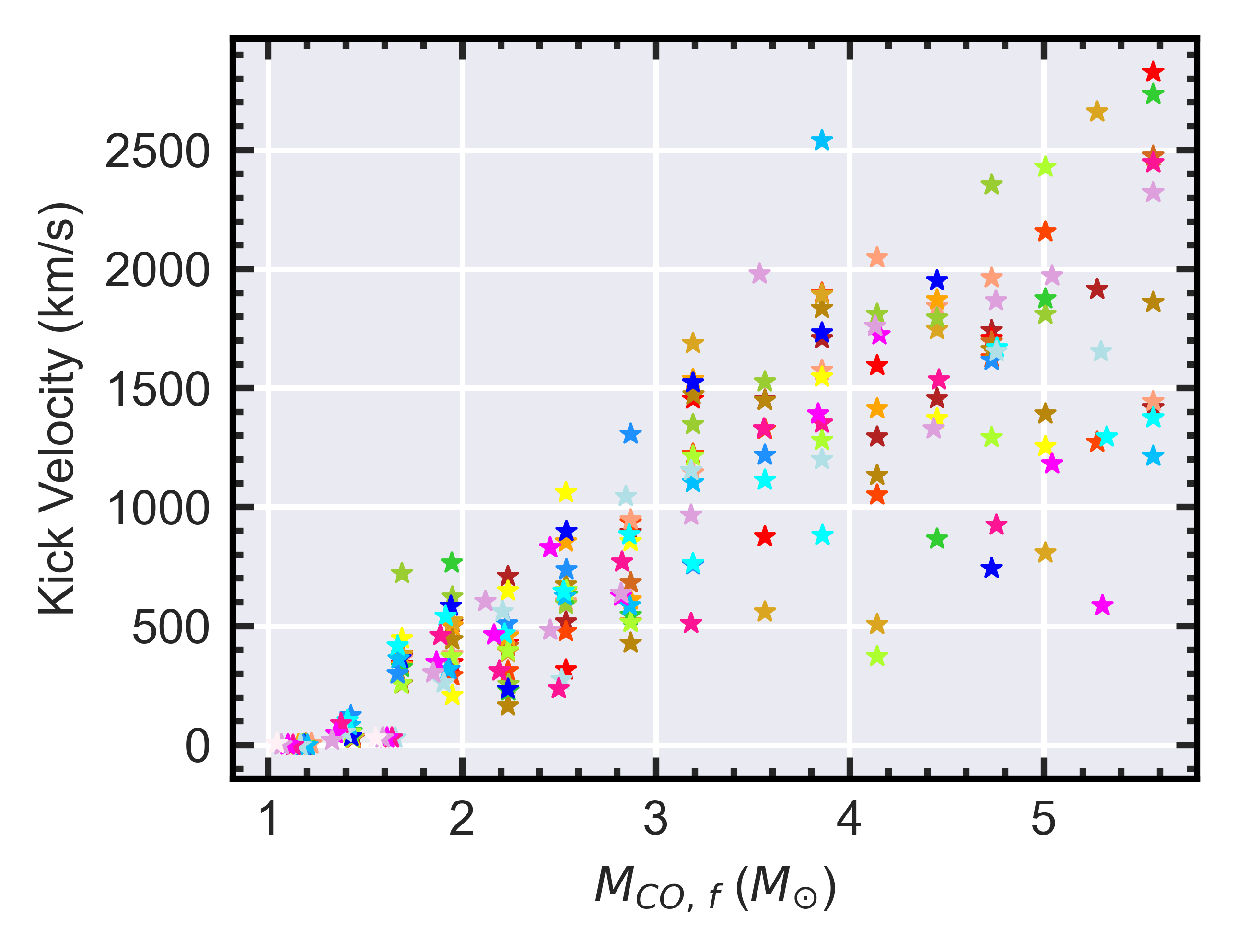}\label{fig:modified_kick_vel}} 
\hspace{0.5cm}

 \caption{Neutron star remnant masses and  natal kick velocity in a binary as a function of CO core masses calculated using the modified model.}
 \label{fig:modified_remnant_mass_kick_vel}
\end{figure*}

\label{modified_model}
\begin{figure*}
\subfigure[\mhe{} = 2.9 \msun{}  ]
{\includegraphics[]{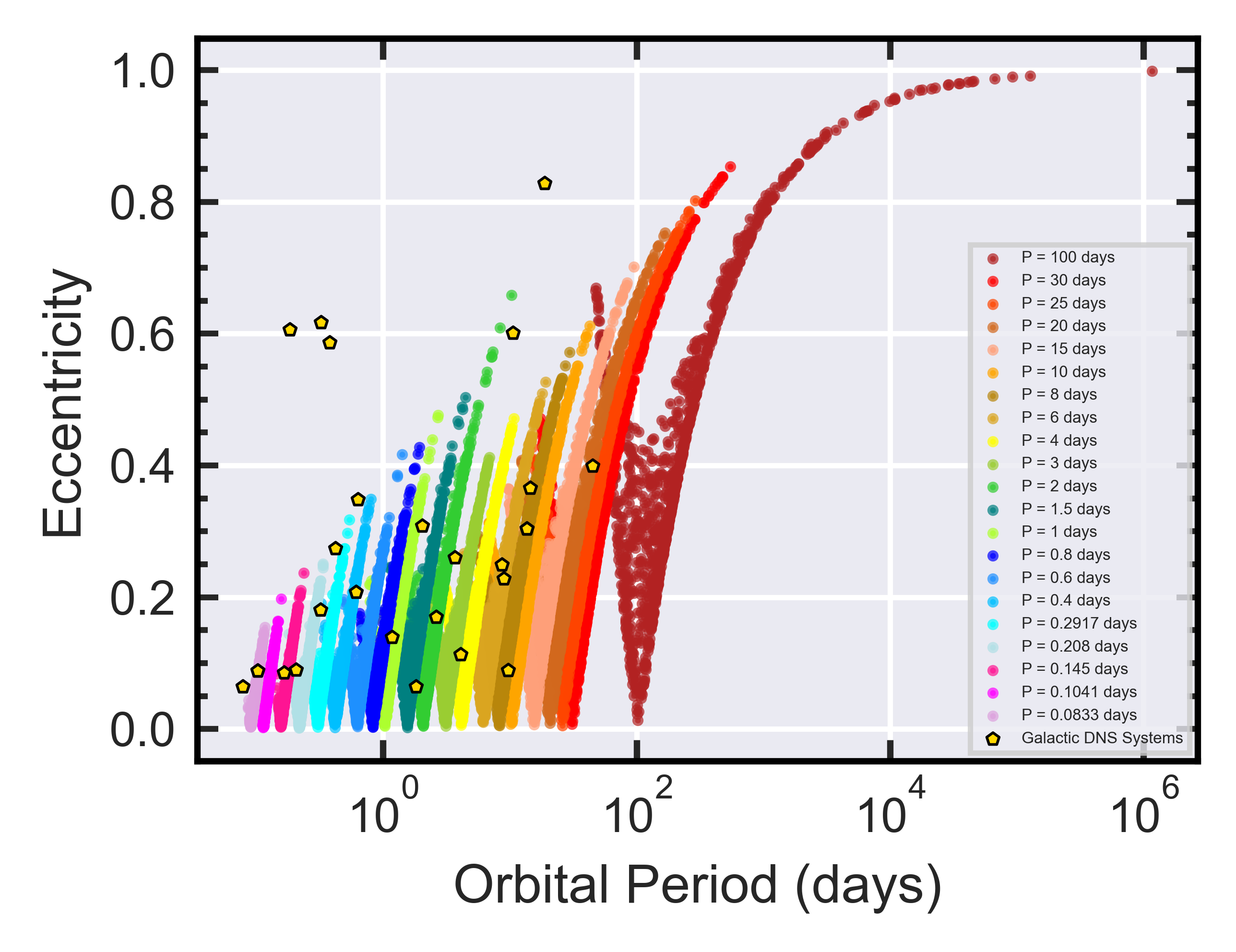}\label{fig:modified_Porb-e_for_2.9M}}
\hspace{0.5cm}
\subfigure[\mhe{} = 3.9 \msun{} ]
{\includegraphics[]{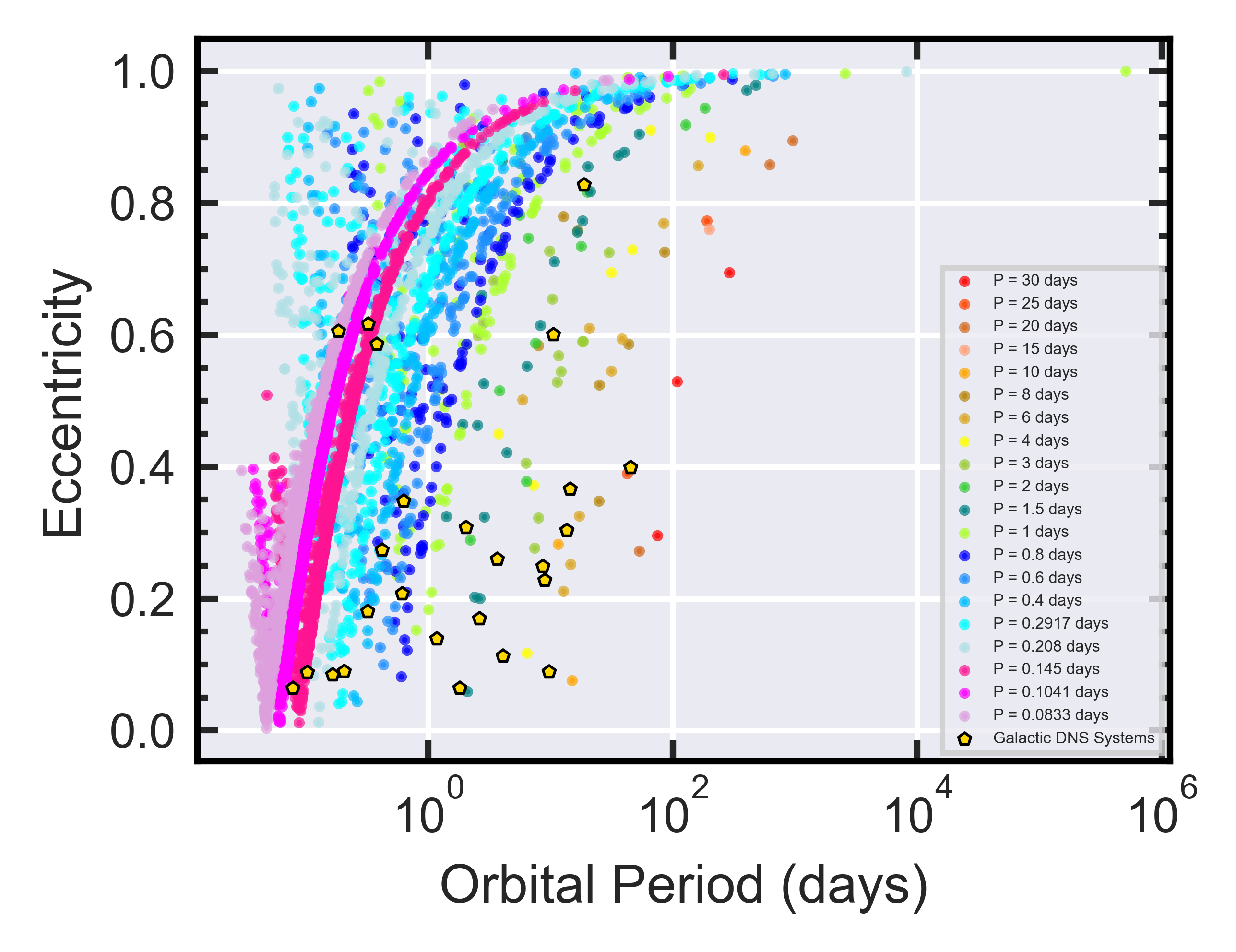}\label{fig:modified_Porb-e_for_4.0M}}
\hspace{0.5cm}
\subfigure[\mhe{} = 5.4 \msun{} ]
{\includegraphics[]{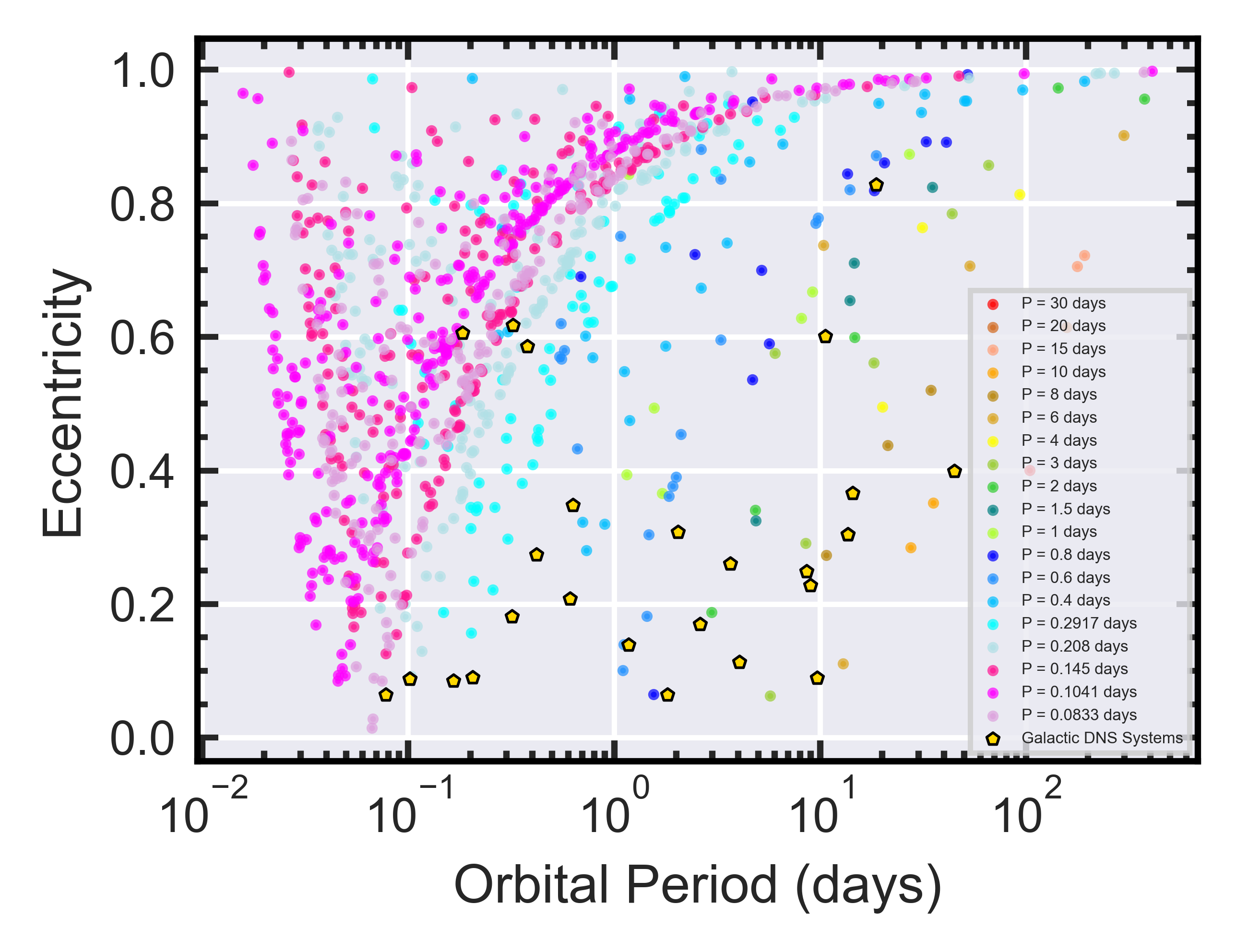}\label{fig:modified_Porb-e_for_5.5M}}
\hspace{0.5cm}
\subfigure[\mhe{} = 8.9 \msun{} ]
{\includegraphics[]{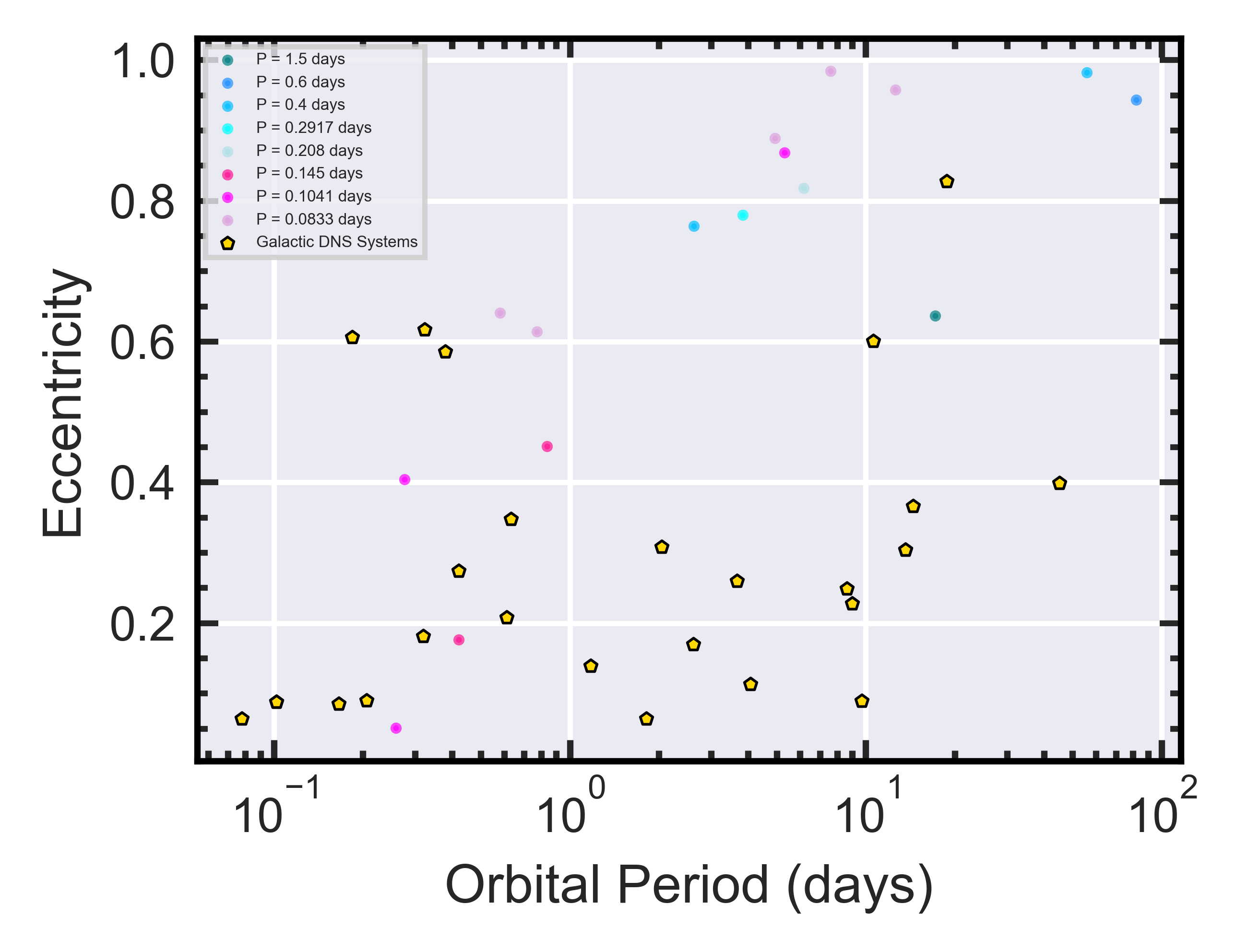}\label{fig:modified_Porb-e_for_8.9M}}
\hspace{0.5cm}

 \caption{Distribution of the simulated post-SN systems in the $P_{\mathrm{orb}}-e$ diagram, using the modified model of SN prescription, in which we performed 1000 cycles with random isotropically oriented NS kicks for each pre-SN system. We adopted pre-SN orbital periods ranging from $0.06$--$100$\,d, which are represented in different colors. }
 \label{fig:porb-e_modified_models}
\end{figure*}

\begin{figure*}
\subfigure[$P_{\mathrm{orb}}-e$ for all the simulated DNS systems]
{\includegraphics[]{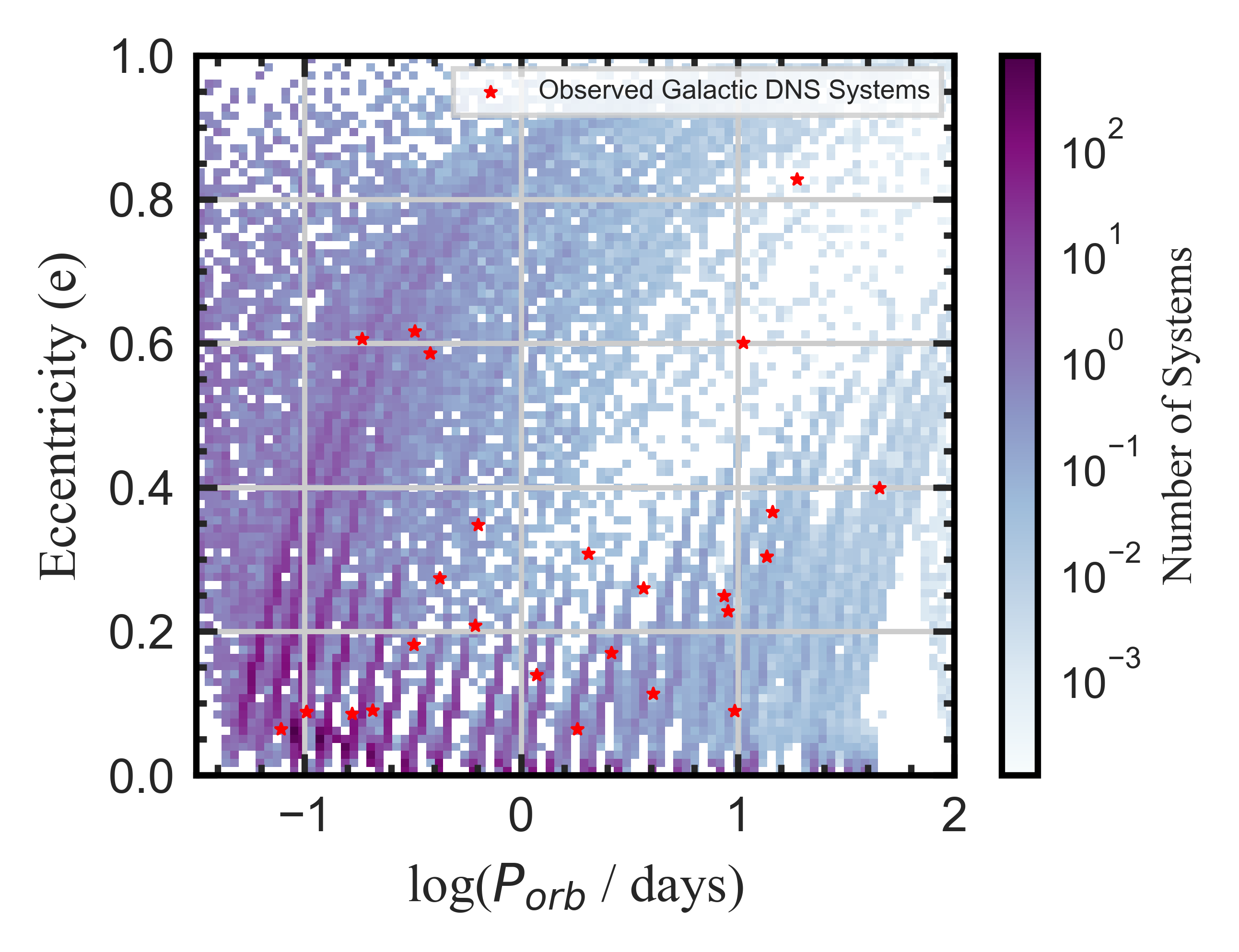}\label{fig:2D_Porb_e_modified}}
\hspace{0.5cm}
\subfigure[$P_{\mathrm{orb}}-e$ for recycled DNS systems]
{\includegraphics[]{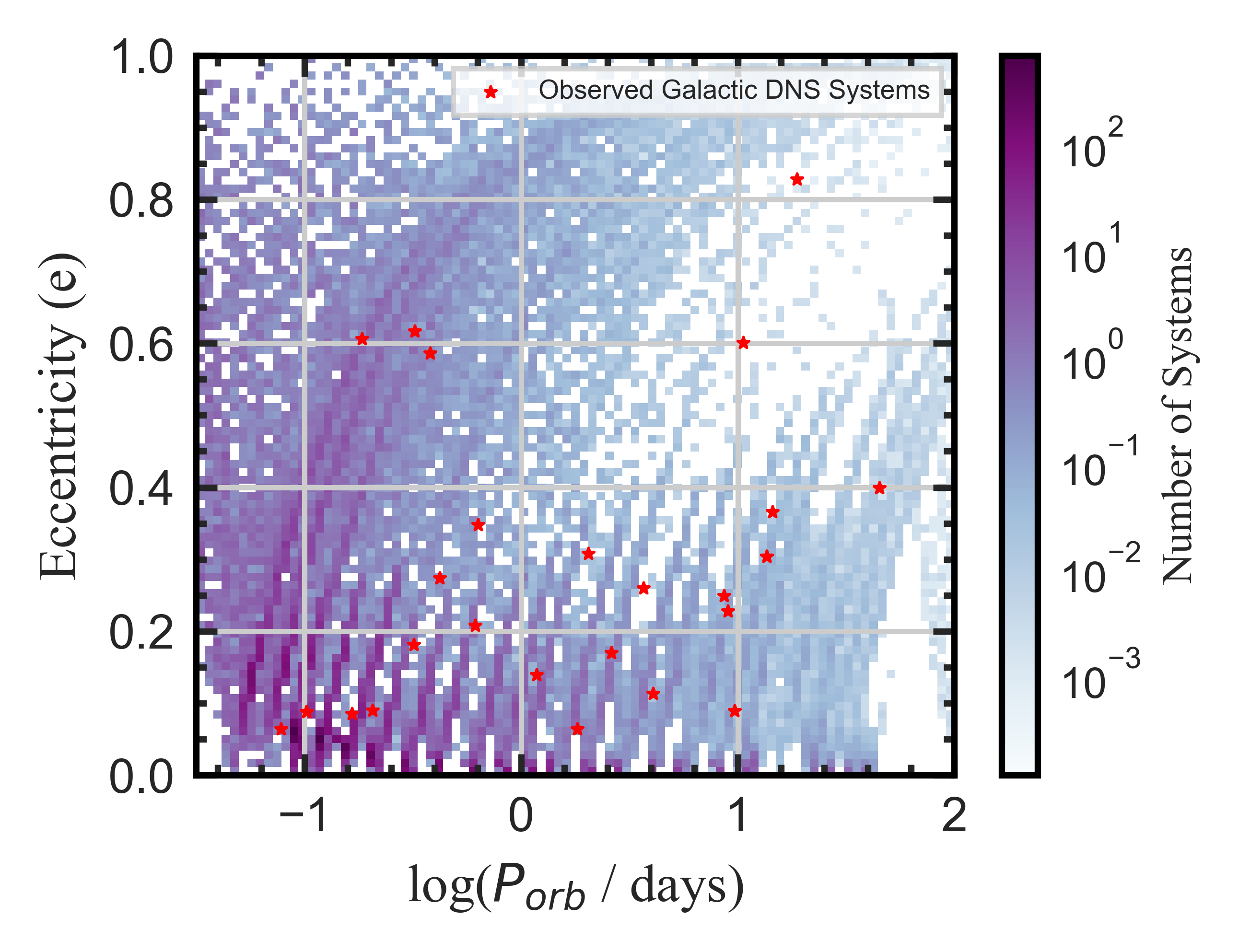}\label{fig:weighted_recycled_2D_Porb_e_modified}}

\hspace{0.5cm}

 \caption{Distribution of all the simulated post-SN systems combined in the $P_\mathrm{orb}$-e diagram (the purple-blue regions) based on the modified model (See text) of SN prescription \citep{Mandel:2020MNRAS, Kapil:2023MNRAS}, in which we set 338 pre-SN P orb that vary logarithmically from $0.06$--$100$\,d. The red stars are the Galactic DNS systems in observations.}
 \label{2D_modified_allsys_and_recycled}
\end{figure*}

Given the discrepancies between the predictions of our standard model and the Galactic DNS distribution, we introduce modifications to the \citet[][]{Mandel:2020MNRAS} remnant prescription. 
We calibrate the models to ensure that they can explain the observed Galactic DNS population. 

The parameter $\mu_{\mathrm{2b}}$ introduced in the \citet[][]{Mandel:2020MNRAS} remnant prescription is based on a linear fit calibrated against a parameterized semi-analytical supernova model \citep{M_ller_2016}. The mass of neutron star remnants is described by a normal distribution $\mathrm{N(\mu, \sigma^2)}$. For progenitor CO core masses in the range M1 $\leq$ \mco{} $\leq$ M2 ($2$\,\msun{} $\leq$ \mco{} $<3$\,\msun{}), the mean neutron star mass is interpolated linearly between two limits:
$\mu = \mu_{\mathrm{2a}} + \mu_{\mathrm{2b}} \cdot (M_{\mathrm{CO}} - M_1)/(M_2 - M_1)$, where $\mu_{\mathrm{2a}}$ sets the base NS mass at \mco{} = M, and $\mu_{\mathrm{2b}}$ defines the slope that governs how the mean NS mass increases with CO core mass in this regime, thereby defining the relationship between the remnant NS masses for CO cores between mass $2$\,\msun{} $\leq$ \mco{} $<3$\,\msun{}.
We vary this free parameter $\mu_{\mathrm{2b}}$ to study its influence on the resulting NS mass distribution and its impact on the post-SN orbital dynamics.
We use $\mu_{\mathrm{2b}}$ = $0.1$ to modify this relation to a more linear one. This avoids the local maxima at CO core mass of $\sim 2.9$\,\msun{} which produces heavy remnant mass of the NSs of $1.8$--$2.0$\,\msun{} (see fig~\ref{fig:modified_remnant masses}). The \citet{Mandel:2020MNRAS} prescription is based on the CO core from single-star evolution, however, the final metal core mass and the final metal core - remnant mass relation can be slightly different for stars in binaries \citep[see][]{Schneider_2020, gilkis2025landscapebinarycorecollapsesupernova}.

We also modify our assumptions for the kick velocities expected from USSN and ECSN. If the model satisfies the condition for USSN as discussed in section~\ref{subsec: SN_kick_prescription}, the kick velocity is reduced by a factor of 3 \citep{Kapil:2023MNRAS}.  
We assume that stars with final CO core masses in the range 1.43--1.65\,\msun{} undergo an ECSN \citep[][]{2003MNRAS.344..629D}. 
This corresponds to ONeMg core masses in the range of 1.37--1.43\,\msun{} \citep{Tauris:2015MNRAS,2015MNRAS.446.2599D, Poelarends2017, 2004ApJ...612.1044P}.
If the model satisfies this condition, the kick velocity is lowered to $30\,\mathrm{kms^{-1}}$. We consider the rule of thumb introduced by \citet{Tauris:2015MNRAS}, that evolutionary tracks where the post-carbon-burning central temperature, $T_c$, rises above the value of $T_c$ $\geq 10^9 K$ during carbon burning will eventually lead to oxygen ignition and later burn silicon to produce an iron core and undergo Fe CCSN. 
For the models in \citet{Tauris:2015MNRAS}, this threshold corresponds to a final ONeMg metal core mass of $1.43$\,\msun{}. 
We map the ONeMg core to the corresponding CO core, as for our models, this is the case for CO core mass $M_\mathrm{CO} > 1.65$\,\msun{}, and we adopt this mass as an approximate threshold limit separating ECSN and Fe CCSN.
Another valuable information that we add is by increasing the kick velocity received by the newborn NS by a factor of 2, for the CO metal core $\geq\,1.66$\,\msun{} as discussed in the Appendix of \citet{Galaudage_2021}. We enable this to kick the slow merging systems visible in radio to higher velocities, simultaneously by making sure that we do not disrupt the GW population.
These modifications give us a better $P_\mathrm{orb}$-e prediction overall. 
See fig.~\ref{fig:modified_remnant_mass_kick_vel} for the modified remnant masses and kick velocities as a function of carbon-oxygen pre-SN core mass. Throughout this paper, we refer to these modifications done to the \citet{Mandel:2020MNRAS} prescription as the \textit{`Modified Model'}.

\begin{center}
    \textbf{A. $P_\mathrm{orb}$-e distribution}
\end{center}
After implementing the modified prescription, the low mass model fig~\ref{fig:modified_Porb-e_for_2.9M} exhibits lower maximum eccentricities compared to the standard model fig~\ref{fig:Porb-e_for_2.9M}. This is a result of the reduced kicks associated with USSN and ECSN in low-mass pre-SN metal cores.
Compared to the standard model, most of the wide orbit systems get disrupted, which is shown in Fig.~\ref {fig:porb-e_modified_models}. On average, approximately only $\sim 18$\,\% of systems survive in the intermediate helium star mass range ($3.5$\,\msun{}-- $5.5$\,\msun{}), leading to the disruption of a large fraction of the systems. Even fewer, $\sim$ $0.7$\,\% double neutron star systems descending from high-mass helium stars ($6.0$--$9.8$\,\msun{}) survive with the modified SN prescription model in comparison to the standard model. 

Fig~\ref{fig:2D_Porb_e_modified} shows the weighted 2D distribution of orbital period ($P_\mathrm{orb}$) versus eccentricity ($e$) for all simulated models post-processed using the modified SN and kick prescription. Three distinct populations are evident. (I) Low eccentricities-short orbital periods, (II) low eccentricities-long orbital periods, (III) high eccentricities-short orbital periods. The low eccentricities - short orbital periods in our models are formed by binaries with low helium stars with kick velocities $\leq 50\,\mathrm{kms^{-1}}$, see fig.~\ref{fig:modified_Porb-e_for_2.9M}. Binaries with a He-star of mass $2.9$\,\msun{} with a few tens of $\mathrm{kms^{-1}}$ of kick velocities can reproduce most of the observed DNS systems in the field except for the DNS systems with eccentricities $\ge 0.6$. 
A few intermediate-mass helium stars with kick velocities of $>100$\,$\mathrm{kms^{-1}}$  can reproduce the tightest observed binaries; however, they do not reproduce wide binaries at low eccentricities (see fig~\ref{fig:modified_Porb-e_for_4.0M}) as most of them get disrupted. 
We note that high-density populations of systems occupying the high eccentricities-short orbital period region of the $P_\mathrm{orb}$--$e$ parameter space typically descend from intermediate-mass helium stars. 
Most of these systems come from the models with a $3.9$\,\msun{} He star with close orbits in the range $0.08$--$0.2$\,d (fig~\ref{fig:modified_Porb-e_for_4.0M}).
Intermediate-mass helium stars with masses $3.4$--$5.4$\,\msun{} with kick velocities exceeding $100$\,$\mathrm{kms^{-1}}$ may form systems like PSR$J1757$-$1854$, PSR$J0509$+$3801$, and PSR$B1534$+$12$. 
This is in agreement with prior studies  \citep{Wang_2006, Wong_2010, Tauris:2017ApJ}.

However, \citet{Andrews:2019ApJ} argue that if DNS systems that lie in this narrow parameter space are the product of isolated binary evolution, the second-born NSs must receive low natal kicks ($< 25$\,$\mathrm{kms^{-1}}$) and descend from He-stars with masses narrowly clustered around $3.2$\,\msun{}. While we can reproduce progenitor models that populate the observed parameter space, such as the $\sim 3.4$\,\msun{} helium star with pre-SN CO core masses of $1.66$\,\msun{}--$1.68$\,\msun{} in orbits of $0.29$--$0.8$ days, we fail to simultaneously account for the high kick velocities ($> 100$ $\mathrm{kms^{-1}}$). This discrepancy may suggest a dynamical origin for such systems, as proposed by \citet{Andrews:2019ApJ}, where close-orbit, high-eccentricity DNS systems form in globular clusters and are subsequently ejected into the Galactic field.

\begin{figure}
    \centering
    \includegraphics[]{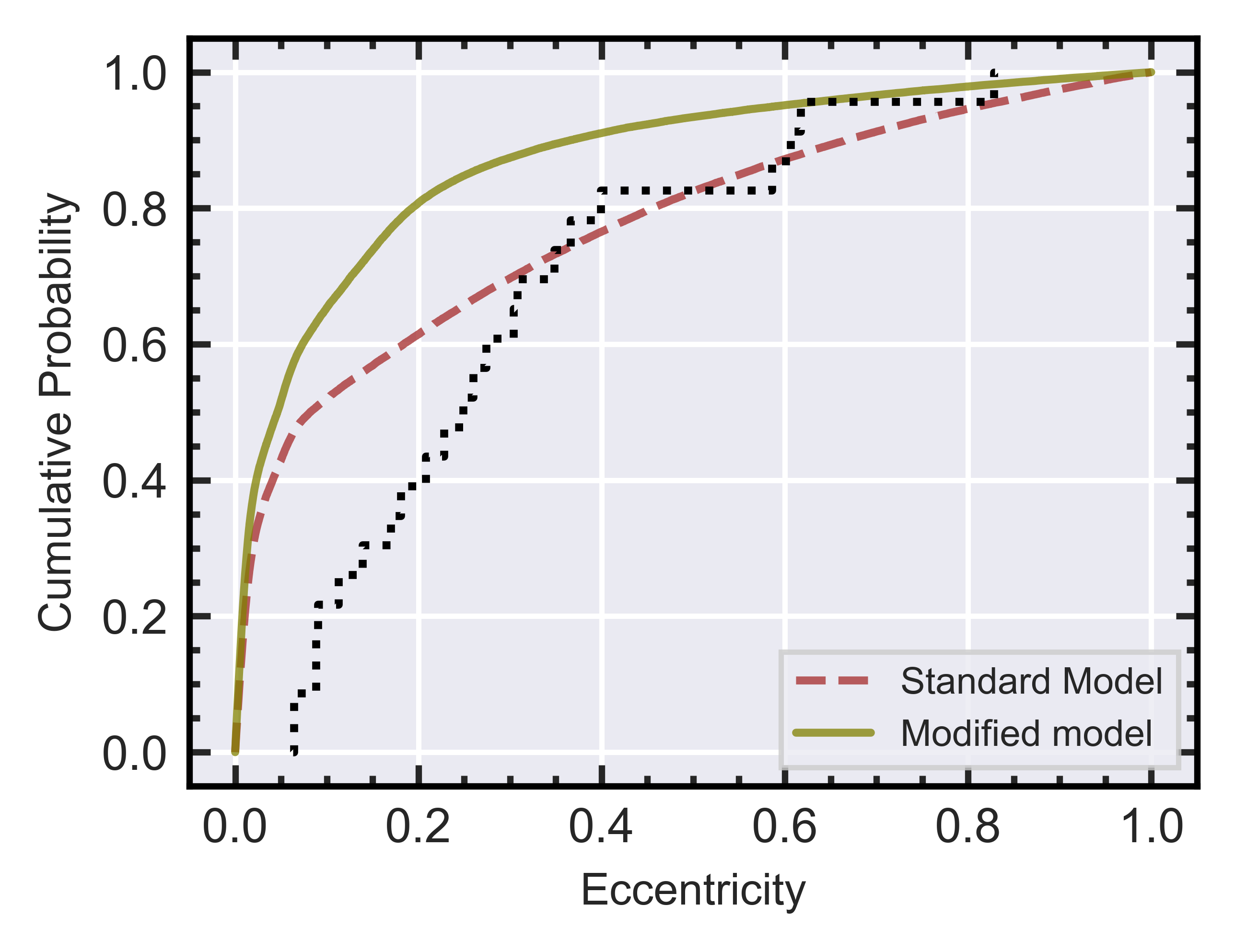}
    \caption{Cumulative distribution for the eccentricity of DNS systems. The dashed red line shows the result from our standard model, whilst the solid yellow line shows the result of the modified model.
    }
    \label{fig:CDF_ecc}
\end{figure}

Although the $P_\mathrm{orb}$--$e$ distribution from the modified prescription shows improved agreement with observations compared to the standard model (see fig.~\ref{fig:CDF_ecc}), it still predicts systems in regions devoid of observed DNSs. 
Notably, three high-density eccentricity lines passing through the PSR\,J1757$-$1854, PSR\,J0509$+$3801, and PSR\,B1534$+$12 systems arise from an intermediate-mass helium star model of $3.9$\,\msun{}. Additionally, a high-density population appears at short orbital periods, across both low and high eccentricities, primarily due to our IMF-based weighting scheme that favors close orbits and low-mass CO cores.

\begin{center}
    \textbf{B. DNS Total mass distribution}
\end{center}

\begin{figure}
    \centering
    \includegraphics[]{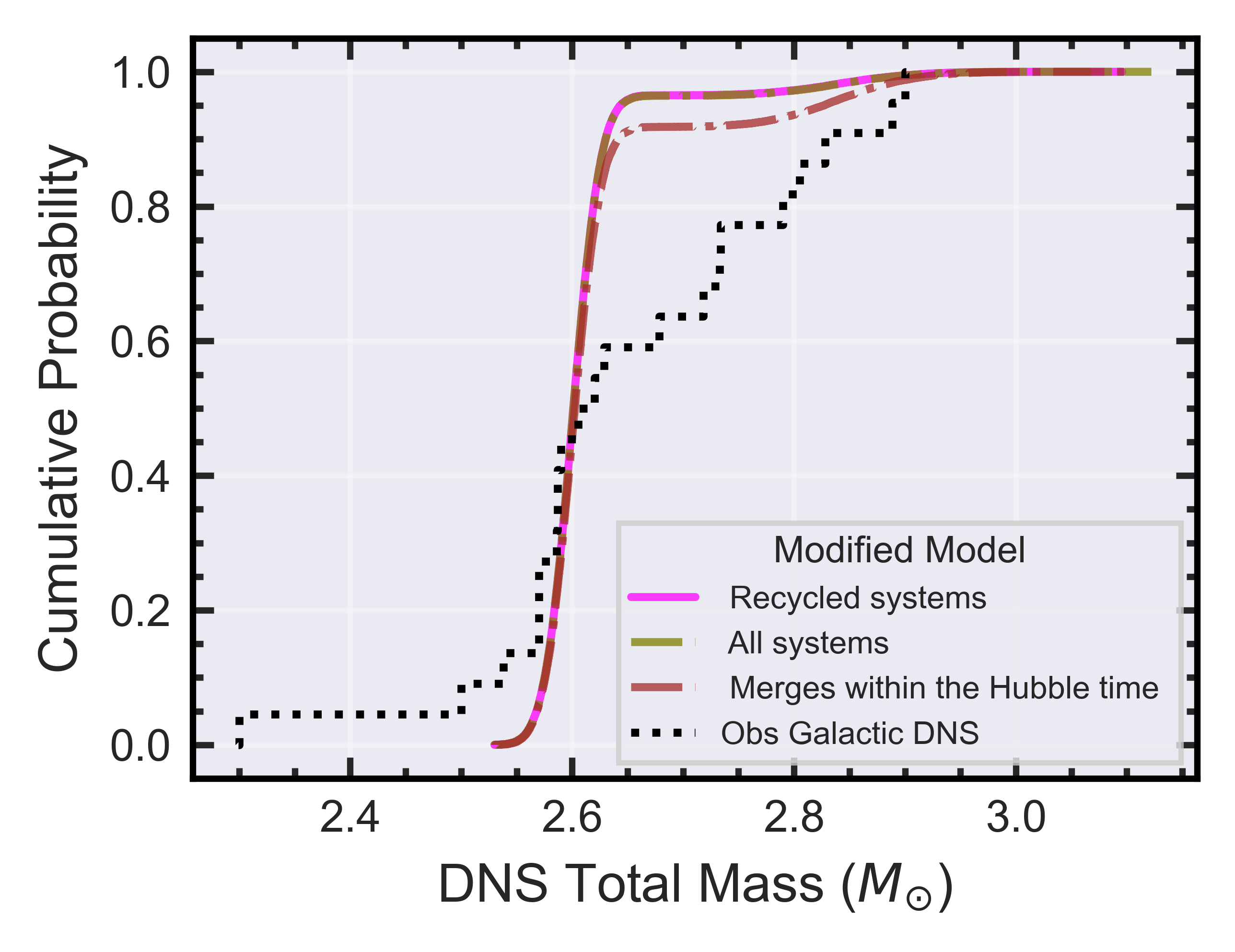}
    \caption{The cumulative distribution of double neutron star total masses. The solid, magenta line represents the recycled systems based on the modified prescription, green dot-dashed line represents all formed double neutron star binaries in the simulation based on the modified prescription. The brown dot-dashed line represents the simulated systems that will merge within the Hubble time, and the black dotted line represents the observed Galactic DNS mass distribution.}

    \label{fig:CDF_Mtot_modified}
\end{figure}

With the modified prescription, our models produces 
DNS systems with total mass ranging from  $2.5$--$3.1$\,\msun{}, whereas the total mass of the observed DNS population in the field ranges from $2.3$--$2.9$\,\msun{}. 
The high-mass tail is reduced here due to the significant changes made to the prescription. This total mass distribution roughly explains the observed DNS total mass distribution.  

Following calibration to ensure consistency with the observed population, the fraction of heavy DNS systems with masses $\ge 3$\,\msun{} in all systems is only $0.01$\,\%.
This small fraction of massive DNSs ($M_\mathrm{tot} \geq 3$\,\msun{}) are mostly produced by binaries with massive helium stars ($5.4$--$9.8$\,\msun{}), which have final CO core masses of $3.4$--$4$\,\msun{}. 
These systems receive kicks as high as $\ge 2000$ $kms^{-1}$, which disrupts the wide binaries. 
Only the compact, high-mass binaries survive. 
The CDF of the total mass distribution of DNS systems is similar to the CDF of the recycled systems, however, $0.02$\,\% of massive systems merge within the Hubble time.

\subsection{GW190425 progenitors}
\label{subsec:GW190425_prog}

Within the subset of massive DNS systems (with total mass $\ge\,3$\,\msun{}) formed in our model, we examine the formation pathways of those that are potential progenitors of GW190425-like systems.
The massive DNS systems are mostly formed by $M_{\mathrm{He, i}} > 7.9$\,\msun{} and a few from systems $\geq 5.5$\,\msun{}. 
Using the standard model, $\sim 5$\,\% of heavy DNS systems are formed in the standard model, and $4.0$\,\% are massive DNSs in the recycled population. 
However,  $0.01$\,\% of all systems consist of heavy systems in the modified model, $0.01$\,\% in the recycled DNS systems, and $0.02$\,\% systems that will merge within Hubble time. This suggests that \textit{the formation of heavy double neutron stars with helium stars at solar metallicity is rare.} 
In our models, within both standard and modified prescriptions, none of the heavy DNS system progenitors undergo unstable mass transfer \citep[][]{Galaudage_2021,2020MNRAS.496L..64R}. However, in the modified model, around $80$\,\% of all the massive DNS systems undergo stable case BB RLO. 
We note that, out of all the heavy DNS systems, around $20$\,\% of massive systems are formed without undergoing any recycling, which results from the massive helium stars with mass $M_{\mathrm{He, i}} > 7.9$\,\msun{}. Larger masses can be indicative of some amount of fallback from the massive helium stars. This can be explained by the formation pathway suggested by \citet{Vigna_G_mez_2021}, which argues that if the massive helium star remains compact and avoids mass transfer, it will avoid recycling the first-born neutron star, and the non-recycled pulsar can become radio quiet and can only be detected in gravitational waves.

\section{Discussion and Conclusions}
\label{sec:conclusions}
     We perform a systematic study of the evolution of helium star-neutron star (He-NS) binaries, which descend from high-mass X-ray (HMXBs) following a common envelope (CE) phase. Successful envelope ejection leaves behind a first-born NS and a stripped OB companion, now a naked helium star. 
     \citet{Yang:2025Science} recently discovered a binary pulsar that may be an example of this evolutionary stage.
     As the helium star expands during shell He burning, it may undergo a second phase of mass transfer via Roche Lobe overflow (case BB RLO) onto the NS. We focus on how the second SN and its associated natal kick influence the post-SN orbital dynamics. Our models are calibrated to reproduce the observed population of Galactic DNS systems in the field. We then explore their implications for GW190425-like systems by comparing different formation channels and assessing which formation scenarios are most likely to produce such massive binaries.

\begin{itemize}
   
    \item We evolved helium stars of masses ranging from $2.5$--$9.8$\,\msun{} from the helium-ZAMS stage until they ignited carbon in a shell. The orbital period corresponding to the initial radius and the maximum radius of the helium star provided us a parameter space to test which systems undergo mass transfer and test the final fate of He - NS binaries. After the binary undergoes Case BB MT, it leaves a pre-SN CO core. We use the SN and kick prescription given by \citep[][]{Mandel:2020MNRAS}, and calibrations from \citet{Kapil:2023MNRAS} to calculate the post-SN remnant masses and kick velocities based on the CO core masses of the helium stars.\\

    \item   
    The standard SN and kick prescription given by \citet[][]{Mandel:2020MNRAS} with calibrations from \citet{Kapil:2023MNRAS} overpredicts the eccentricities (e) between $0.3$--$0.6$ and $P_\mathrm{orb}$ ranging from a few hours to $1$\,d corresponding to eccentricities $> 0.4$, as shown in fig.~\ref{fig:std_model_2d_porb-e}. This is because \citet[][]{Mandel:2020MNRAS} overpredicts the explosion energies of ECSN and follows the same kick prescription as other NS natal kicks and the kick mechanism used for CCSN, which overall contributes to a major uncertainty associated with the kick velocities associated with newborn NSs. Additionally, the total mass of our simulated systems ranges from $2.5$--$3.4$\,\msun{}, whereas the observed population of DNSs in the field ranges from $2.3$--$2.9$\,\msun{}. The fraction of DNSs with total mass $\ge3.0$\,\msun{} within the recycled population of our simulated models is $4.09$\,\%. 
    This contribution stems from the assumed relationship between the CO core masses and remnant masses, which produces a local maxima at CO core mass of $2.9$\,\msun{}, which produces NSs of masses ranging between $1.6$--$2.0$\,\msun{}. Until now, no massive DNS system has been observed in the Galactic field. This can indicate one of two things: firstly, that a better understanding of pre-SN metal core and remnant mass relation, kicks imparted onto the newborn NS, depending on the particular SN explosion is needed, or secondly, it could mean that massive systems might exist in the radio population, only that we have not observed one.\\
    
    \item To account for these issues and calibrate the models to the observed population, we modified the standard prescription by imposing three conditions: (1) scaling down the NS remnant mass parameter $\mu_{\mathrm{2b}}$ which is responsible for determining the remnant masses resulting from CO cores ranging between $2$\,\msun{} $\leq$ \mco{} $< 3$\,\msun{}. Our modified models results in a more linear relation between the CO core mass and post-SN remnant masses. (2) assuming lower kick velocities of a few tens of $\mathrm{kms^{-1}}$ for stars that explode as USSN and ECSN, and (3) kicking the wide binaries to higher velocities. Our modified model still produces a population of close orbit $P_\mathrm{orb} < 1$\,d, high eccentricity $0.3 >$ e $> 0.6$ DNSs in the region of $P_\mathrm{orb}$-e parameter space where no radio DNS candidate has been observed yet. 
    These systems come mainly from intermediate-mass helium stars ($3.4$\msun{} $<$ \mhe{} $< 5.5$\,\msun{}). Perhaps accounting for the final pre-SN metal mass range for ECSN correctly can solve the eccentricities that appear in the observed gap. The only way to self-consistently account for the variation in the ONeMg core that accurately decides whether the star will undergo ECSN would be to run our models to the end of carbon burning/onset of core-collapse \citep[cf.][]{Jiang:2021ApJL}. 
    Recently \citet{Valli:2025arXiv} identified two distinct types of supernova explosions in Be X-ray binaries, one producing neutron stars with ultralow kicks ($< 10$\,$\mathrm{kms^{-1}}$) and another with polar-aligned kicks clustered around $100$\,$\mathrm{kms^{-1}}$. They show that the two new kick modes can also explain the observed DNS $P_\mathrm{orb}$-e population, which appears to split into two distinct groups. By applying the same kick properties derived from Be X-ray binaries, the models successfully reproduce the two observed populations in the $P_\mathrm{orb}$-e plane of DNS systems. This fit supports scenarios with lower ejecta masses, consistent with the known DNS progenitor expectations. A detailed exploration of how incorporating this information affects our model is beyond the scope of the present study and will be addressed in future work.\\

    \item In terms of DNS total mass distribution, there is not any significant difference between the mass distribution of all systems and the recycled systems. We do not see any DNS system with total mass $< 2.51$\,\msun{} because the smallest metal core in our model produces neutron stars of remnant masses $\geq 1.2$\,\msun{}, which results in DNS total masses $> 2.5$\,\msun{}.
    The high-mass tail is reduced in the modified model due to adjustments in the CO core–remnant mass relationship, specifically setting $\mu_{\mathrm{2b}} = 0.1$.
    Massive DNS systems with total masses $M_\mathrm{tot} > 3$\,\msun{} are mostly produced by binaries with massive helium stars with masses of $5.4$--$9.8$\,\msun{}, which have final CO core masses of $3.4$--$4$\,\msun{}. 
    
    These systems recieve kicks as high as $\sim 2000$\,km/s, which disrupt most wide binaries, such that only the compact, high-mass binaries remain bound.
    Using the modified prescription, within all the systems and the recycled population, only produces $0.01\%$ of heavy double neutron stars, which is roughly in agreement with the observed population. However, with the current study, where we assume binary models with helium stars at solar metallicity, we are unable to explain the high rate of GW190425. Out of all the massive DNSs $\ge 3$\,\msun{}, around $80$\,\% of the progenitors of massive DNS systems undergo stable case BB RLO MT, whilst approximately $20$\,\% avoid mass transfer completely \citep[cf.][]{Vigna_G_mez_2021}. Altogether, our models rule out the formation of massive DNSs like GW190425 via unstable MT at \textit{solar metallicity} \citep[][]{2020MNRAS.496L..64R, Abbott:2020ApJL, Galaudage_2021}.\\ 
    
This study is limited to models assuming a fixed first-born neutron star mass of $1.4$\,\msun{} and modeling He–NS binary evolution at solar metallicity. These simplifications constrain our ability to explore additional uncertainties relevant to the formation of GW190425, as outlined in Section~\ref{sec:intro}. In Part II of this work, we will systematically examine how the evolution of double neutron star systems is influenced by lower metallicities, assess how different first-born neutron star masses, particularly more massive neutron stars as proposed by \citet[][]{Qin:2024A&A}, affect binary outcomes, and investigate the consequences of allowing for super-Eddington accretion \citep[cf.][]{Zhang:2023MNRAS}.

Having calibrated our simulations to reproduce the observed double neutron star population, this extended framework will provide a foundation to systematically test key uncertainties in the formation of GW190425-like systems and establish a self-consistent, unbiased formation pathway for massive DNS systems such as GW190425.

\end{itemize}

\section*{Acknowledgements}

We thank the referee, Yong Shao, for comments that improved the manuscript.
This research was supported by the Australian Research Council (ARC) Centre of Excellence for Gravitational Wave Discovery (OzGrav), through project numbers CE170100004 and CE230100016. 
SS and AN are supported by an ARC Discovery Early Career Research Award (DE220100241; PI Stevenson). 
This work was performed on the OzSTAR national facility at Swinburne University of Technology. 
The OzSTAR program receives funding in part from the Astronomy National Collaborative Research Infrastructure Strategy (NCRIS) allocation provided by the Australian Government, and from the Victorian Higher Education State Investment Fund (VHESIF) provided by the Victorian Government.

\section*{Data availability}

MESA version: \texttt{23.05.01} inlists for He-NS binary evolution and 
postprocessing scripts to simulate DNS formation can be found here: \href{https://github.com/nair-ashwathi/BinaryNeutronStar_Formation}{GitHub repository}.




\bibliographystyle{mnras}
\bibliography{bib} 






\bsp	
\label{lastpage}
\end{document}